\documentclass{article}

\usepackage[main,final]{neurips_2026}

\usepackage[utf8]{inputenc} 
\usepackage[T1]{fontenc}    
\usepackage{hyperref}       
\usepackage{url}            
\usepackage{booktabs}       
\usepackage{amsfonts}       
\usepackage{nicefrac}       
\usepackage{microtype}      
\usepackage{xcolor}         

\definecolor{darkblue}{rgb}{0.0, 0.17, 0.58}
\definecolor{darkgreen}{rgb}{0.0, 0.5, 0.0}
\usepackage{tabularx}
\usepackage{graphicx}
\usepackage{booktabs}
\usepackage{multirow}
\usepackage{amsmath} 
\usepackage{graphicx}
\usepackage{enumitem}
\usepackage{fancybox} 
\usepackage{subcaption}
\usepackage{caption}
\usepackage{makecell}
\usepackage{wrapfig}
\usepackage[most]{tcolorbox}
\usepackage{xcolor}
\usepackage{listings}  
\usepackage{float}
\usepackage{hyperref}
\usepackage{colortbl}
\usepackage{array}   
\usepackage{amssymb}
\usepackage{tikz}
\usepackage{wrapfig}
\usepackage{titletoc}
\usepackage{adjustbox}
\usepackage{placeins}
\usepackage{needspace}
\definecolor{darkblue}{rgb}{0.0, 0.17, 0.58}
\hypersetup{colorlinks,breaklinks,linkcolor=red,citecolor=darkblue}

\newcolumntype{L}[1]{>{\raggedright\arraybackslash}p{#1}}
\newcolumntype{Y}{>{\raggedright\arraybackslash}X}

\newcommand{\fulldashline}{%
\\[-0.35em]
\multicolumn{2}{@{}c@{}}{%
\begin{tikzpicture}
\draw[gray!45, line width=0.35pt, dash pattern=on 4pt off 4pt] (0,0) -- (\linewidth,0);
\end{tikzpicture}%
}
\\[-0.35em]
}

\newcommand{\tabindent}{\hspace*{0.7em}}

\title{NINJA: A Navigator–Inspector Joint Architecture for Context-Efficient Issue Localization}

\author{%
  \textbf{Wenkai Fang\textsuperscript{1,2} \enspace
  Shunyu Liu\textsuperscript{3}\thanks{Corresponding author.} \enspace
  Wei Luo\textsuperscript{2} \enspace
  Zihao Wu\textsuperscript{1} \enspace
  Yang Zhou\textsuperscript{1} \enspace
  Tongya Zheng\textsuperscript{4}} \\
  \textbf{Wotao Yin\textsuperscript{2} \enspace
  Jialie Shen\textsuperscript{5} \enspace
  Leszek Rutkowski\textsuperscript{6} \enspace
  Mingli Song\textsuperscript{1} \enspace
  Dacheng Tao\textsuperscript{3}} \\
  \textsuperscript{1}Zhejiang University \quad
  \textsuperscript{2}DAMO Academy, Alibaba Group \\
  \textsuperscript{3}College of Computing and Data Science,
  Nanyang Technological University, Singapore \\
  \textsuperscript{4}Hangzhou City University \\
  \textsuperscript{5}City St George's, University of London \\
  \textsuperscript{6}AGH University of Science and Technology, Poland \\
  \texttt{wenkfang@zju.edu.cn} \quad \texttt{shunyu.liu.cs@gmail.com}
}

\begin{document}

\maketitle

\begin{abstract}
Recent advances in agent-based methods have demonstrated strong promise for issue localization, a critical prerequisite for software issue resolution.
However, most existing agent-based methods rely on a single agent with a growing context, where long-context accumulation compresses the effective reasoning space. Meanwhile, the flat exploration structure hinders the balance between file-level breadth and function-level depth.
To address these limitations, we propose NINJA, a hierarchical \textit{\textbf{N}avigator-\textbf{IN}spector \textbf{J}oint \textbf{A}rchitecture} for context-efficient issue localization. NINJA decomposes repository exploration into global navigation and local inspection: the navigator maintains the global search state, performs file-level search, and dispatches selected entry files, while inspectors independently conduct function-level exploration around the assigned files in separate contexts. Through multi-round interactions, inspectors return suspicious locations as feedback, and the navigator updates the global state to decide whether to continue exploration or finalize localization. This hierarchical design balances file-level breadth with function-level depth, while independent inspector contexts prevent local exploration traces from accumulating in a single growing context. To further strengthen both global coordination and local exploration, we introduce a two-stage agentic fine-tuning strategy.
Extensive experiments across multiple benchmarks and LLM backbones show that NINJA consistently outperforms competitive baselines. Notably, after fine-tuning, Qwen3-Coder-30B-A3B-Instruct surpasses the strong closed-source Claude-Haiku-4.5 model. 
Our code is available at \url{https://github.com/wantbook-book/NINJA}.
\end{abstract}

\section{Introduction}

Large language model (LLM) agents have recently achieved substantial progress across diverse domains, including software engineering~\citep{jiang2025agentic,liu2024large,xia2025live}, web navigation~\citep{zheng2025deepresearcher,jin2025search}, and research~\citep{hou2026paperdebugger,song2026paperorchestra,tu2026spagent,zhou2025breaking}. In software engineering, LLM agents have shown promising results on issue localization tasks~\citep{chen2025locagent,jiang2025cosil}. Issue localization is a key prerequisite for repository-level issue resolution, since resolving an issue typically requires identifying issue-relevant code locations before applying a fix~\citep{yang2024sweagent,xia2024agentless,xie2025swefixer}. Accurate localization can substantially improve downstream resolve rates, as supported by prior studies~\citep{jiang2025cosil,ma2025tool,he2026sweadept,sutawika2026codescout} and our downstream results in Sec.~\ref{sec:downstream_utility}.

Early issue localization methods mainly retrieve code locations semantically relevant to a natural-language issue description~\citep{suresh2024cornstack,reddy2025swerank}, with reflective retrieval improving this paradigm through query refinement and re-ranking~\citep{reddy2025swerank+}. Yet these methods largely rely on static code representations and struggle to capture multi-hop structural relations needed for root-cause localization~\citep{chen2025locagent,liu2025graphlocator,fehr2025coret}. 
Agent-based issue localization offers a more flexible alternative: agents interact with repositories through tools such as file reading, code search, and dependency inspection, while iteratively revising localization decisions based on new evidence~\citep{qin2024agentfl,xia2024agentless,huang2024understanding,yao2022react}.

Nevertheless, most agent-based issue localization methods still rely on a single agent that explores the repository within one continuously growing context. 
As earlier evidence and intermediate search results accumulate, past trajectories increasingly consume the context budget and cause long-context accumulation. This compresses the effective reasoning space for later-stage localization decisions~\citep{cui2025agentic} and makes subsequent exploration decisions more difficult.
Moreover, issue localization is inherently a complex exploration task~\citep{chen2025locagent}, requiring both broad repository-level search to identify potentially relevant files and deep local reasoning to trace caller--callee relations within specific file regions~\citep{yu2025orcaloca}. 
However, a single-agent framework follows a flat and sequential exploration structure, requiring the same agent to handle breadth expansion, depth inspection, and their budget allocation within one trajectory. This lack of hierarchical decomposition places a heavy burden on single-agent decision making and makes balanced exploration difficult.

To address these limitations, we propose NINJA, a \textit{\textbf{N}avigator-\textbf{IN}spector \textbf{J}oint \textbf{A}rchitecture} that uses a hierarchical division of labor to avoid single-context accumulation and balance broad repository search with deep local inspection.
Technically, NINJA coordinates a global navigator with dynamically scheduled inspectors through multi-round interactions. The navigator performs file-level search and dispatches promising entry files, while each inspector explores the assigned file region in an independent context, without inheriting the navigator's accumulated trajectory. The navigator then uses the feedback of inspectors to update its global search state and decide whether to continue exploration or finalize localization.
We further train NINJA in two stages: after SFT warmup for both components, we sequentially apply component-wise RL, first to the inspector and then to the navigator. Our key contributions are summarized as follows:

\begin{itemize}[leftmargin=*]

\item We propose NINJA, a hierarchical navigator--inspector joint architecture for issue localization. Through multi-round navigator--inspector interactions, NINJA balances file-level breadth with function-level depth, while decomposing exploration into independent file-region inspections to keep each inspector's context within a roughly constant budget.

\item We introduce a two-stage agentic fine-tuning for NINJA. After SFT warmup for both components, we sequentially optimize the inspectors and the navigator with component-wise RL, improving local function-level exploration, global search coordination, and their multi-round interaction.

\item Extensive experiments demonstrate that NINJA achieves superior localization performance over strong baselines. With two-stage agentic fine-tuning, the open-source Qwen3-Coder-30B-A3B-Instruct further surpasses the strong closed-source Claude-Haiku-4.5 model.

\end{itemize}

\section{Related work}
\subsection{Issue localization}

%

Issue localization is a critical prerequisite for software maintenance and program repair, aiming to identify issue-relevant code locations from a natural-language issue description~\citep{chen2025locagent}. Early methods formulate this task as semantic retrieval between issues and code~\citep{suresh2024cornstack,reddy2025swerank}. More recently, LLM-based agents have emerged for repository-level localization, using graph or code indices~\citep{chen2025locagent,liu2025graphlocator}, coarse-to-fine search from files or classes to functions and methods~\citep{yu2025orcaloca,jiang2025cosil}, or SFT/RL to improve localization policies~\citep{ma2025tool,zhang2025one}. Procedure-based methods further constrain agent execution with hand-crafted workflows while retaining LLM-based code reasoning~\citep{xia2024agentless,liu2025graphlocator,wang2025improving,jiang2025cosil}. However, most agent-based methods still rely on a single growing context, where accumulated trajectories consume the context budget and compress later-stage reasoning space~\citep{cui2025agentic}. Although context pruning and relevance filtering can remove irrelevant information~\citep{yu2025orcaloca,jiang2025cosil}, they cannot avoid accumulation across multiple relevant file regions as repositories scale. 
To address this, we decompose issue localization into global navigation and independent local inspections: the navigator coordinates file-level search, global state updates, and multi-round dispatch, while inspectors explore assigned file regions in independent contexts. This hierarchy reduces single-context accumulation and balances file-level breadth with function-level depth.

\subsection{Agentic fine-tuning}

Recent LLM advances increasingly rely on post-training to improve instruction following~\citep{zhang2025replay,liu2025survey}, reasoning~\citep{guo2025deepseekr1,fang2025serl,zhang2025consistent,zhang2026simple}, and task-specific alignment~\citep{liu2025mathematical,tu2025global,liu2024odyssey}. A common recipe first uses SFT to teach desired formats and behaviors, followed by RL with task-specific rewards~\citep{yang2025qwen3,liu2024deepseek,team2026qwen35}. However, agentic tasks extend beyond single-turn generation, requiring models to make multi-step decisions, invoke tools, and revise subsequent actions based on environment feedback~\citep{jin2025search,zheng2025deepresearcher,lv2024codeact}. This has motivated agentic post-training methods and resources, including trajectory-based SFT for learning interaction patterns~\citep{zeng2024agenttuning,chen2024agentflan,chen2025locagent}, outcome-based RL for optimizing task success~\citep{ma2025tool,zhang2025one}, efficiency-oriented training strategies that focus on critical decision steps~\citep{chen2025atlas}, and large-scale agent datasets for training and evaluation~\citep{yang2025swe,song2024agentbank}. Following this line of work, we use \emph{agentic fine-tuning} to refer to post-training with multi-step interaction trajectories. In our work, we apply agentic fine-tuning to NINJA to improve the navigator's global coordination, the inspector's local code inspection, and their multi-round interaction during issue localization.
\section{Background}
\label{sec:backgrounds}
\subsection{Problem formulation}


An issue localization agent aims to identify repository code locations relevant to a given issue description. Each location is represented as $(f_i, u_j)$, where $f_i$ denotes a file and $u_j$ denotes the corresponding function or class method, such as $(\texttt{file1.py}, \texttt{func1})$ or $(\texttt{file1.py}, \texttt{class1.method1})$. We denote the resulting issue-relevant location set as $\mathcal{G}$. Existing issue localization benchmarks typically define ground truth based on edited locations, namely the code locations modified in the resolving PR patch, rather than all locations that may be relevant for resolving the issue. 
Following prior work~\citep{chen2025locagent}, we extract edited locations from the resolving PR patch as $\mathcal{G}_e^\star$.

We adopt ReAct~\citep{yao2022react} as the agent execution paradigm. Given an initial query $q$ consisting of the issue description and task instruction, the agent interacts with the repository through iterative reasoning, action, and observation steps. At step $t$, the agent receives the current context
\begin{equation}
c_t = (q, r_1, a_1, o_1, \dots, r_{t-1}, a_{t-1}, o_{t-1}),
\end{equation}
where $r_i$, $a_i$, and $o_i$ denote the reasoning trace, action, and observation at step $i$, respectively. The agent then generates a reasoning trace $r_t$ and an action $a_t$, where the action space includes both tool calls and the terminal action $\texttt{exit}$. Since reasoning and action are produced by the same LLM via autoregressive decoding, we model them with a unified policy:
\begin{equation}
r_t \sim \pi_\theta(\cdot \mid c_t), \qquad
a_t \sim \pi_\theta(\cdot \mid c_t, r_t).
\end{equation}
After executing $a_t$, the environment returns an observation $o_t$, which is appended to the context for the next step.
The interaction terminates when the agent selects $\texttt{exit}$, when the number of steps exceeds $T_{\max}$, or when the token consumption exceeds $B_{\max}$. Upon termination, we construct the final context by concatenating the current context $c_t$ with a finalization prompt $q_{\text{final}}$. The model then produces the final predicted edited-location set:
\begin{equation}
c_{t+1} = c_t \oplus [q_{\text{final}}], \qquad
\hat{\mathcal{G}}_e \sim \pi_\theta(\cdot \mid c_{t+1}).
\end{equation}

\subsection{Agentic fine-tuning}

We fine-tune the agent in two stages. We first apply SFT as a warmup stage to teach the model the required output format, tool-use protocol, and basic exploration behavior. We then apply RL with verified rewards to further optimize the agent's long-horizon search decisions.

\textbf{SFT warmup.}
Let $\mathcal{D}_{\mathrm{sft}}=\{(q_i,\tau_i)\}_{i=1}^{N}$ denote the SFT dataset, where $q_i$ is the initial query consisting of the issue description and task instruction, and $\tau_i$ is the target ReAct trajectory, including reasoning traces, tool calls, observations, and the final edited-location output. We define an assistant-only mask $b_{i,m}\in\{0,1\}$, which equals 1 for assistant-generated tokens and 0 otherwise. Tool observations and externally supplied prompts are retained as conditioning context but excluded from the loss. Following the unified policy $\pi_\theta$ defined above, we optimize the masked next-token prediction objective, normalized by the number of assistant tokens in each trajectory:
\begin{equation}
\mathcal{L}_{\mathrm{SFT}}(\theta)
=
-\frac{1}{N}
\sum_{i=1}^{N}
\frac{1}{\sum_{m=1}^{|\tau_i|} b_{i,m}}
\sum_{m=1}^{|\tau_i|}
b_{i,m}\log \pi_{\theta}\left(\tau_{i,m}\mid q_i,\tau_{i,<m}\right).
\end{equation}
This stage initializes the agent with basic ReAct-style exploration behavior and reduces tool-execution failures and format-following errors before reinforcement learning.

\textbf{Reinforcement learning.}
Starting from the SFT-warmed policy $\pi_\theta$, we further optimize the agent with PPO~\citep{schulman2017proximal}. 
For each query $q$, the old policy $\pi_{\theta_{\mathrm{old}}}$ samples a trajectory $\tau$ through interactions with the available tools, ultimately producing a predicted edited-location set $\hat{\mathcal{G}}_e$.
A verifier assigns a reward $R(\tau)$ by comparing $\hat{\mathcal{G}}_e$ with the ground truth $\mathcal{G}_e^\star$, providing a task-level learning signal for localization accuracy.
Let $\mathcal{I}_{\mathrm{asst}}(\tau)$ denote the positions of assistant-generated tokens in $\tau$. The clipped PPO policy loss is computed only over these positions:
\begin{equation}
\mathcal{L}_{\mathrm{PPO}}^{\pi}(\theta)
=
-\mathbb{E}_{\tau,\,m\in\mathcal{I}_{\mathrm{asst}}(\tau)}
\left[
\min
\left(
\rho_m(\theta)\hat{A}_m,
\operatorname{clip}\left(\rho_m(\theta),1-\epsilon,1+\epsilon\right)\hat{A}_m
\right)
\right],
\end{equation}
where $\epsilon$ specifies the clipping coefficient, and $\rho_m(\theta)$ denotes the importance sampling ratio for an assistant-generated token $\tau_m$ at position $m\in\mathcal{I}_{\mathrm{asst}}(\tau)$:
$
\rho_m(\theta)
=
\frac{
\pi_{\theta}(\tau_m \mid q,\tau_{<m})
}{
\pi_{\theta_{\mathrm{old}}}(\tau_m \mid q,\tau_{<m})
}.
$
The advantage $\hat{A}_m$ is estimated using generalized advantage estimation (GAE)~\citep{schulman2015high} with a critic $V_\phi$, based on the verified localization reward.

\begin{figure}[!t]
    \centering
    \includegraphics[width=\linewidth]{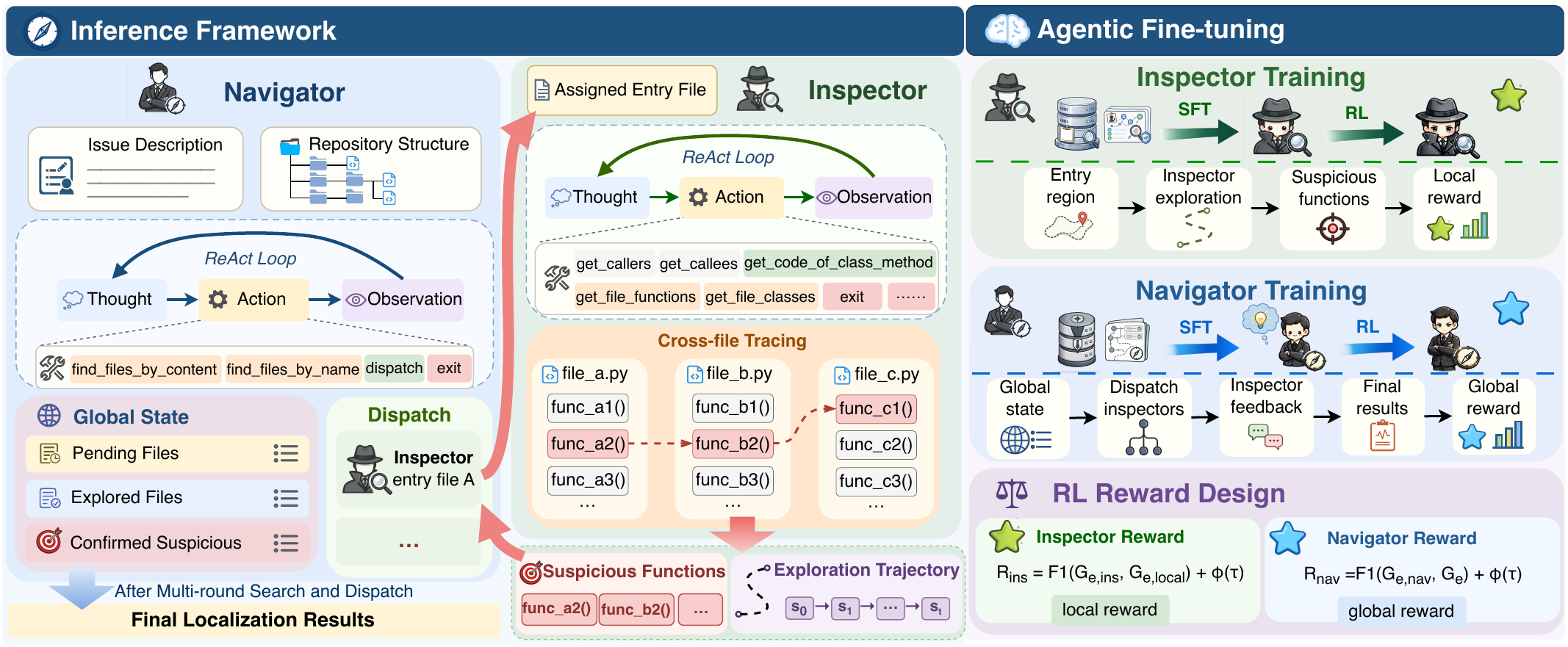}

    \caption{Overview of the NINJA framework and agentic fine-tuning pipeline. (1) NINJA decomposes repository exploration into global navigation and local inspection, where the navigator searches and dispatches issue-relevant files to inspectors, updates the global state from their returned findings, and iteratively continues exploration or finalizes the results. (2) The fine-tuning pipeline combines SFT warmup with component-wise RL, training the inspector before the navigator.}
    
    \label{fig:framework}
    \vspace{-0.3cm}
\end{figure}

\section{Method}
To achieve high-performance and context-efficient issue localization, we propose NINJA, a navigator--inspector joint architecture with two-stage agentic fine-tuning. NINJA separates file-level coordination from function-level inspection, reducing single-context history accumulation and enabling multi-round exploration that balances file-level breadth with function-level depth. We further use SFT warmup to teach the required output format and increase tool-call execution success, followed by component-wise RL to improve navigator coordination and inspector localization.

\subsection{NINJA framework}
As shown in Fig.~\ref{fig:framework}, NINJA consists of a navigator and multiple inspectors. The navigator maintains a repository-level search state, performs file-level search, and dispatches promising entry files, while inspectors independently conduct function-level exploration around assigned files. This division of labor reduces single-context accumulation and balances file-level breadth with function-level depth.

\textbf{Navigator.}
The navigator serves as the global coordinator of NINJA. Given the issue description and the repository directory structure, it maintains a global search state
$
\mathcal{Z}_t = (\mathcal{S}_t, \mathcal{E}_t, \mathcal{Q}_t)
$
at step~$t$, where $\mathcal{S}_t$ is the suspicious function set, $\mathcal{E}_t$ is the explored file set, and $\mathcal{Q}_t$ is the pending file queue. Specifically, $\mathcal{S}_t$ stores candidate issue-relevant functions or methods returned by inspectors, $\mathcal{E}_t$ records files that have already been investigated to avoid redundant dispatch, and $\mathcal{Q}_t$ contains candidate files waiting for further inspection.
At each step, the navigator reviews the current global state $\mathcal{Z}_t$ and decides the next exploration action. If the existing evidence is sufficient, it terminates the search and outputs the final localization result. Otherwise, it either searches the repository to identify additional candidate files or dispatches inspectors to files in $\mathcal{Q}_t$ using the navigator tools listed in Appendix~\ref{app:tools}.

\textbf{Inspector.}
Each inspector independently starts from an entry file dispatched by the navigator, using only the issue description and assigned entry file without inheriting the navigator's accumulated trajectory.
It inspects functions and class methods using the tools listed in Appendix~\ref{app:tools}, selectively retrieves implementations that appear issue-relevant, and reasons about whether they may contribute to the reported behavior. The inspector is not limited to the entry file: when a suspicious function depends on external logic, it can trace callers and callees to better understand the surrounding issue-relevant code region and identify the underlying root cause. This allows the inspector to move across files when necessary while keeping exploration focused around the local execution context. Once no further relevant functions are worth exploring, it returns the identified suspicious functions with supporting explanations to the navigator, along with a record of visited files and functions.

\textbf{Navigator--inspector interaction.}
When the navigator invokes the \texttt{dispatch} tool at step $t$, it selects a set of entry files $\mathcal{U}_t \subseteq \mathcal{Q}_t$ with $|\mathcal{U}_t|=M_t$ for further inspection. Since the parallelism size $P$ is a tunable hyperparameter, at most $P$ inspectors can run simultaneously. Therefore, $\mathcal{U}_t$ is partitioned into $L_t=\lceil M_t/P\rceil$ mini-batches, denoted as $\{\mathcal{C}_{t,\ell}\}_{\ell=1}^{L_t}$, such that
\begin{equation}
\mathcal{U}_t = \bigcup_{\ell=1}^{L_t} \mathcal{C}_{t,\ell}, 
\quad 
\mathcal{C}_{t,\ell} \cap \mathcal{C}_{t,\ell'} = \emptyset \ \text{for } \ell \neq \ell',
\quad
|\mathcal{C}_{t,\ell}| \leq P.
\end{equation}
Each mini-batch $\mathcal{C}_{t,\ell}$ is processed in parallel by at most $P$ inspectors, while successive mini-batches are executed sequentially to incorporate feedback from earlier exploration.
We use $\mathcal{Z}_{t,\ell}=(\mathcal{S}_{t,\ell},\mathcal{E}_{t,\ell},\mathcal{Q}_{t,\ell})$ to denote the temporary state after the $\ell$-th mini-batch, with $\mathcal{Z}_{t,0}=\mathcal{Z}_t$.
Within each mini-batch, inspectors independently explore their assigned entry files, returning suspicious functions with supporting explanations as well as records of visited files and functions.
After each mini-batch, the navigator updates the temporary state from $\mathcal{Z}_{t,\ell-1}$ to $\mathcal{Z}_{t,\ell}$.
Specifically, it adds returned suspicious functions to $\mathcal{S}_{t,\ell}$, records dispatched files plus any additional files visited by inspectors in $\mathcal{E}_{t,\ell}$, and removes these files from $\mathcal{Q}_{t,\ell}$.
Entry files already visited by earlier inspectors are skipped in subsequent mini-batches, reducing redundant exploration while preserving parallelism within each mini-batch.
Once all mini-batches are processed, the dispatch call completes, yielding the next navigator state $\mathcal{Z}_{t+1}=\mathcal{Z}_{t,L_t}$.
Detailed prompts are provided in Appendix~\ref{app:prompts}.

\subsection{Agentic fine-tuning}

To improve tool use, exploration, and coordination, we fine-tune NINJA in two stages. First, we use Claude-Haiku-4.5 to generate teacher trajectories, filter out invalid-format and incorrect-localization samples, and use the remaining trajectories for SFT warmup. This stage teaches the open-source LLM the required output format, tool-use protocol, and basic exploration patterns. Second, we apply component-wise RL to improve multi-step exploration: we first train the inspector for stronger local inspection, and then train the navigator with the optimized inspector fixed as its feedback provider.


\textbf{Reward for the navigator.}
Following RepoNavigator~\citep{zhang2025one}, we define the navigator's reward as the weighted sum of a set-level localization score and a tool-execution success score:
\begin{equation}
\mathcal{R}_{\mathrm{nav}}(\hat{\mathcal{G}}_{e}, \mathcal{G}^{\star}_{e}, \tau_{\mathrm{nav}})
=
\mathrm{F1}(\hat{\mathcal{G}}_{e}, \mathcal{G}^{\star}_{e})
+
\lambda_{\phi}\phi(\tau_{\mathrm{nav}}),
\end{equation}
where $\hat{\mathcal{G}}_{e}$ is the parsed and normalized predicted edited-location set, $\mathcal{G}^{\star}_{e}$ is the ground-truth edited-location set, and $\tau_{\mathrm{nav}}$ is the navigator trajectory. The term $\phi(\tau_{\mathrm{nav}})\in[0,1]$ measures the execution success rate of tool calls, and $\lambda_{\phi}$ controls its weight. We set $\lambda_{\phi}=1$ in our experiments.
For set-level matching, we use the F1 score:
\begin{equation}
\label{eq:f1}
\mathrm{F1}(\mathcal{A}, \mathcal{B})
=
\frac{2|\mathcal{A}\cap \mathcal{B}|}
{|\mathcal{A}|+|\mathcal{B}|},
\end{equation}
where $\mathrm{F1}(\mathcal{A}, \mathcal{B})\in[0,1]$. A tool call is considered unsuccessful if it violates the required format, refers to a non-existent symbol, or terminates abnormally during execution.

\textbf{Reward for the inspector.}
Unlike the navigator, which performs global coordination, the inspector is evaluated only against the ground-truth locations within the scope of its own exploration.

Let $\mathcal{F}(\tau_{\mathrm{ins}})$ denote the set of files visited by an inspector trajectory $\tau_{\mathrm{ins}}$. We define the locally scoped ground-truth edited-location set as
\begin{equation}
\mathcal{G}^{\star}_{e,\mathrm{local}}(\tau_{\mathrm{ins}})
=
\left\{
g=(f_i,u_j) \in \mathcal{G}^{\star}_{e}
\;\middle|\;
f_i \in \mathcal{F}(\tau_{\mathrm{ins}})
\right\}.
\end{equation}

Let $\hat{\mathcal{G}}_{e,\mathrm{ins}}$ denote the parsed and normalized suspicious edited-location set returned by the inspector. The inspector's reward is defined as
\begin{equation}
\mathcal{R}_{\mathrm{ins}}(\hat{\mathcal{G}}_{e,\mathrm{ins}}, \mathcal{G}^{\star}_{e,\mathrm{local}}(\tau_{\mathrm{ins}}), \tau_{\mathrm{ins}})
=
\mathrm{F1}(\hat{\mathcal{G}}_{e,\mathrm{ins}}, \mathcal{G}^{\star}_{e,\mathrm{local}}(\tau_{\mathrm{ins}}))
+
\lambda_{\phi}\phi(\tau_{\mathrm{ins}}).
\end{equation}
When $\mathcal{G}^{\star}_{e,\mathrm{local}}(\tau_{\mathrm{ins}})=\emptyset$, the trajectory is excluded from training because it does not provide an effective supervision signal for local localization.

Overall, the component-wise reward design aligns the inspector with local function-level exploration and the navigator with global coordination, while encouraging successful tool use for both agents.

\section{Experiments}
\label{sec:experiments}
\subsection{Experimental setups}
\textbf{Training settings.}
Following prior work~\citep{zhang2025one}, we use SWE-smith-py~\citep{yang2025swe} for both SFT and RL, with Qwen3-Coder-30B-A3B-Instruct and Qwen3-14B~\citep{yang2025qwen3} as base models. To prevent data leakage, we apply repository-level filtering to ensure that no repository in the training set appears in the evaluation benchmarks, regardless of whether the corresponding issues differ. Detailed training settings are provided in Appendix~\ref{app:train_settings}.

\textbf{Evaluation settings.} We evaluate our method and the baselines using two recent models, Claude-Haiku-4.5~\citep{anthropic2025haiku45} and Qwen3-Coder-30B-A3B-Instruct, on two issue localization benchmarks: SWE-Bench-Lite~\citep{jimenez2023swe} and LocBench~\citep{chen2025locagent}. 
Following prior work~\citep{xu2026learning,chen2025locagent,zhang2025one}, we measure localization performance using Acc@$k$ and $\overline{\mathrm{F1}}$, and report token cost and runtime to characterize computational overhead. For token quantities, k and K denote multipliers of 1,000 and 1,024, respectively. Details of the evaluation metrics and hyperparameters are provided in Appendix~\ref{app:evaluation_metrics}.

\textbf{Baselines.} We compare NINJA with representative issue localization baselines, including retrieval-based CodeRankEmbed~\citep{suresh2024cornstack}, retrieval-enhanced SWERank+~\citep{reddy2025swerank+}, graph-based LocAgent~\citep{chen2025locagent}, hierarchical CoSIL~\citep{jiang2025cosil}, action-scheduling OrcaLoca~\citep{yu2025orcaloca}, and lightweight toolset design RepoSearcher~\citep{ma2025tool}. Implementation details are provided in Appendix~\ref{app:baselines_comparison}.

\begingroup

\begin{table*}[t]
\small
\centering
\setlength{\tabcolsep}{6pt}
\renewcommand{\arraystretch}{1.15}

\newcolumntype{C}{>{\centering\arraybackslash}X}
\newcolumntype{P}[1]{>{\raggedright\arraybackslash}p{#1}}

\caption{Results on SWE-Bench-Lite. Best and second-best localization results are in \textbf{bold} and \underline{underlined}, respectively. CodeRank, Haiku-4.5, and Qwen3-30B denote CodeRankEmbed, Claude-Haiku-4.5, and Qwen3-Coder-30B-A3B-Instruct. NINJA-14B and NINJA-30B are obtained by applying agentic fine-tuning (SFT+PPO) to Qwen3-14B and Qwen3-30B, respectively.}
\label{tab:main_results_swe_bench_lite}

\begin{tabularx}{\textwidth}{P{1.6cm}P{1.55cm}CCCCCC}
\toprule
\multirow{2}{*}{Framework}
  & \multirow{2}{*}{Model}
  & \multicolumn{2}{c}{File-level}
  & \multicolumn{2}{c}{Function-level}
  & \multirow{2}{*}{\# Toks (k)}
  & \multirow{2}{*}{Time (s)} \\
\cmidrule(lr){3-4}\cmidrule(lr){5-6}
  &
  & Acc@5
  & F1 
  & Acc@5
  & F1
  &
  & \\
\midrule

Embedding
  & CodeRank
  & $75.0{\pm}0.2$
  & $28.4{\pm}0.1$
  & $54.0{\pm}0.2$
  & $19.6{\pm}0.1$
  & N/A
  & $14.9{\pm}0.1$ \\

\midrule

\rowcolor[HTML]{EDEEEE}
  & Haiku-4.5
  & $45.6{\pm}0.0$
  & $20.3{\pm}0.0$
  & $17.8{\pm}0.1$
  & $8.0{\pm}0.1$
  & $41.0{\pm}0.1$
  & $150.2{\pm}0.5$ \\
\rowcolor[HTML]{EDEEEE}
\multirow{-2}{*}{SWERank+}
  & Qwen3-30B
  & $26.8{\pm}0.6$
  & $8.9{\pm}0.2$
  & $1.1{\pm}0.0$
  & $0.4{\pm}0.0$
  & $53.2{\pm}0.0$
  & $140.0{\pm}0.7$ \\

\multirow{2}{*}{LocAgent}
  & Haiku-4.5
  & $83.1{\pm}0.4$
  & $31.0{\pm}0.2$
  & $37.2{\pm}0.4$
  & $13.2{\pm}0.1$
  & $37.8{\pm}0.2$
  & $128.6{\pm}0.5$ \\
  & Qwen3-30B
  & $44.0{\pm}3.1$
  & $14.7{\pm}1.0$
  & $25.4{\pm}3.0$
  & $9.3{\pm}1.2$
  & $46.7{\pm}1.8$
  & $52.6{\pm}0.3$ \\

\rowcolor[HTML]{EDEEEE}
  & Haiku-4.5
  & $82.3{\pm}0.3$
  & $31.6{\pm}0.2$
  & $68.3{\pm}0.4$
  & $26.6{\pm}0.1$
  & $25.7{\pm}0.1$
  & $75.7{\pm}0.3$ \\
\rowcolor[HTML]{EDEEEE}
\multirow{-2}{*}{CoSIL}
  & Qwen3-30B
  & $79.2{\pm}0.3$
  & $29.8{\pm}0.1$
  & $24.6{\pm}0.4$
  & $12.9{\pm}0.2$
  & $24.3{\pm}0.0$
  & $33.3{\pm}0.1$ \\

\multirow{2}{*}{OrcaLoca}
  & Haiku-4.5
  & $70.7{\pm}0.3$
  & $\underline{57.6{\pm}0.2}$
  & $52.0{\pm}0.4$
  & $29.2{\pm}0.1$
  & $26.7{\pm}0.1$
  & $135.0{\pm}0.5$ \\
  & Qwen3-30B
  & $64.0{\pm}0.5$
  & $51.8{\pm}0.5$
  & $39.0{\pm}0.7$
  & $27.9{\pm}0.2$
  & $30.5{\pm}0.2$
  & $121.7{\pm}1.3$ \\

\rowcolor[HTML]{EDEEEE}
  & Haiku-4.5
  & $81.7{\pm}0.4$
  & $28.5{\pm}0.1$
  & $29.0{\pm}0.4$
  & $27.3{\pm}0.2$
  & $23.3{\pm}0.1$
  & $102.1{\pm}0.4$ \\
\rowcolor[HTML]{EDEEEE}
\multirow{-2}{*}{RepoSearcher}
  & Qwen3-30B
  & $74.8{\pm}0.5$
  & $26.0{\pm}0.1$
  & $19.9{\pm}1.5$
  & $13.9{\pm}0.5$
  & $26.7{\pm}0.0$
  & $61.3{\pm}0.3$ \\

\midrule

\multirow{5}{*}{NINJA (Ours)}
  & Haiku-4.5
  & $\underline{84.0{\pm}0.4}$
  & $\mathbf{59.7{\pm}0.4}$
  & $\underline{76.3{\pm}0.5}$
  & $\underline{29.3{\pm}0.2}$
  & $36.5{\pm}0.2$
  & $154.1{\pm}0.6$ \\
  & Qwen3-14B
  & $72.0{\pm}1.0$
  & $47.9{\pm}0.8$
  & $50.3{\pm}1.0$
  & $19.8{\pm}0.2$
  & $35.0{\pm}0.1$
  & $57.4{\pm}0.8$ \\
  & NINJA-14B
  & $83.3{\pm}0.4$
  & $49.7{\pm}0.3$
  & $51.8{\pm}0.4$
  & $20.3{\pm}0.2$
  & $32.0{\pm}0.2$
  & $45.7{\pm}0.3$ \\
  & Qwen3-30B
  & $79.3{\pm}0.7$
  & $53.5{\pm}0.7$
  & $71.1{\pm}0.4$
  & $27.7{\pm}0.6$
  & $51.7{\pm}0.5$
  & $85.4{\pm}3.2$ \\
  & NINJA-30B
  & $\mathbf{90.6{\pm}0.4}$
  & $\mathbf{59.7{\pm}0.5}$
  & $\mathbf{77.3{\pm}0.5}$
  & $\mathbf{29.4{\pm}0.6}$
  & $48.7{\pm}0.4$
  & $84.7{\pm}3.0$ \\

\bottomrule
\end{tabularx}

\end{table*}

\endgroup
\subsection{Main results}

As shown in Tab.~\ref{tab:main_results_swe_bench_lite} and the additional results in Appendix~\ref{app:locbench_results}, NINJA achieves strong overall performance. For localization performance, NINJA generally outperforms existing baselines across both backbone models. Notably, with our agentic fine-tuning strategy, Qwen3-Coder-30B-A3B-Instruct further surpasses Claude-Haiku-4.5, demonstrating the effectiveness of our training approach. 
NINJA's performance gains come with additional runtime. Nevertheless, across the agent-based baselines in Tab.~\ref{tab:main_results_swe_bench_lite}, its runtime remains within roughly \(2\times\) in most cases and at most \(2.6\times\). Before agentic fine-tuning, NINJA with Qwen3-30B is also faster than SWERank+ and OrcaLoca.

\begingroup

\begin{table*}[t]
\small
\centering
\setlength{\tabcolsep}{6pt}
\renewcommand{\arraystretch}{1.15}

\newcolumntype{C}{>{\centering\arraybackslash}X}

\caption{Ablation study of our NINJA framework on SWE-Bench-Lite.}
\label{tab:ablation}

\begin{tabularx}{\textwidth}{
  >{\raggedright\arraybackslash}p{3.5cm}
  CCCCCC
}
\toprule
\multirow{2}{*}{Framework}
  & \multicolumn{2}{c}{File-level}
  & \multicolumn{2}{c}{Function-level}
  & \multirow{2}{*}{\# Toks (k)}
  & \multirow{2}{*}{Time (s)} \\
\cmidrule(lr){2-3}\cmidrule(lr){4-5}
  &
  Acc@5
  & $\overline{\mathrm{F1}}$
  & Acc@5
  & $\overline{\mathrm{F1}}$
  &
  & \\
\midrule

NINJA
  & $79.3{\pm}0.7$
  & $53.5{\pm}0.7$
  & $71.1{\pm}0.4$
  & $27.7{\pm}0.6$
  & $51.7{\pm}0.5$
  & $85.4{\pm}3.2$ \\

\tabindent w/o independent context
  & $43.5{\pm}0.6$
  & $31.4{\pm}0.7$
  & $41.5{\pm}0.3$
  & $18.1{\pm}0.5$
  & $24.6{\pm}0.3$
  & $17.2{\pm}0.4$ \\

\tabindent w/o multi-round dispatch
  & $69.9{\pm}0.6$
  & $48.7{\pm}0.6$
  & $65.6{\pm}0.4$
  & $23.5{\pm}0.6$
  & $41.7{\pm}0.5$
  & $63.4{\pm}1.9$ \\
  
\tabindent w/o dynamic scheduling
  & $77.3{\pm}0.6$
  & $51.7{\pm}0.7$
  & $69.9{\pm}0.4$
  & $26.8{\pm}0.6$
  & $58.7{\pm}0.6$
  & $93.8{\pm}2.7$ \\

\bottomrule
\end{tabularx}
\vspace{-0.4cm}
\end{table*}

\endgroup
\subsection{Ablation study}
\textbf{Settings.}
To investigate the sources of NINJA's performance gains and characterize its computational trade-offs, we conduct an ablation study on SWE-Bench-Lite using Qwen3-Coder-30B-A3B-Instruct as the backbone model. (1) We first consider three ablated variants of NINJA. 
(i) \textit{w/o independent context} merges the navigator and inspectors into a single agent, which performs both file-level and function-level search within the same context. 
(ii) \textit{w/o multi-round dispatch} requires the navigator to finalize localization immediately after the first \texttt{dispatch} call returns, disabling further rounds of file search and inspector dispatch. 
(iii) \textit{w/o dynamic scheduling} removes the navigator's global tracking of explored files, making inspectors more likely to be redundantly dispatched to files that have already been visited during previous cross-file exploration.
(2) To rule out alternative explanations for the performance gains, we additionally conduct controlled experiments on token budget and task-specific fine-tuning. We evaluate NINJA under aggregate token budgets comparable to those of existing methods and apply the same SFT+PPO procedure to the single-agent OrcaLoca framework.
(3) We vary the degree of inspector parallelism $P$ to analyze its effect on localization performance and computational cost. This setting is important because parallelism affects not only exploration latency but also dynamic scheduling behavior: larger parallel batches lead to less frequent updates of the pending file queue, which may increase duplicated exploration.

\textbf{Sources of performance gains.}
As shown in Tab.~\ref{tab:ablation}, 
(1) \textit{w/o independent context} leads to a substantial performance drop.
When file-level search and function-level inspection share a single context, repository-level exploration quickly accumulates long histories, which compress the model's effective reasoning space and introduce redundant information, making subsequent localization decisions more difficult.
In contrast, NINJA decomposes localization into multiple independent contexts with a clear division of labor: the navigator maintains the global state, dispatches pending files to inspectors for function-level inspection, and summarizes the final results once sufficient evidence is gathered. This design allows each agent to focus on a narrower subtask in a separate context, reducing long-context compression and redundant history accumulation, and thus enables more reliable localization.
(2)~The \textit{w/o multi-round dispatch} variant also degrades performance. This suggests that a single round of exploration is often insufficient for complex issues, where the issue description may only expose surface-level symptoms~\citep{reddy2025swerank+}. Multiple rounds of dispatch enable the navigator to progressively integrate evidence, refine its understanding of the bug, and identify the root cause.

\textbf{Token-budget and fine-tuning controls.}
We further examine whether NINJA's improvements can be attributed to greater token consumption or task-specific training. Under token budgets matched to CoSIL and OrcaLoca (Tab.~\ref{tab:budget_control}), NINJA achieves substantially higher function-level Acc@5 while maintaining competitive performance across other localization metrics. To control for training effects, we apply the same SFT+PPO pipeline to OrcaLoca (Tab.~\ref{tab:finetuning_control}). Although fine-tuning improves OrcaLoca's performance, NINJA maintains a clear advantage both before and after training. Together, these results suggest that NINJA's gains primarily stem from its framework design rather than increased token consumption or task-specific training.

\begingroup

\begin{table*}[t]
\small
\centering
\setlength{\tabcolsep}{6pt}
\renewcommand{\arraystretch}{1.15}

\newcolumntype{C}{>{\centering\arraybackslash}X}

\caption{Token-matched comparison on SWE-Bench-Lite using Qwen3-Coder-30B-A3B-Instruct. NINJA limits the aggregate trajectory tokens of the navigator and all inspectors per task, and finalizes localization once the budget is exhausted.}

\label{tab:budget_control}

\begin{tabularx}{\textwidth}{
  >{\raggedright\arraybackslash}p{3.4cm}
  CCCCCC
}
\toprule
\multirow{2}{*}{Setting}
  & \multicolumn{2}{c}{File-level}
  & \multicolumn{2}{c}{Function-level}
  & \multirow{2}{*}{\# Toks (k)}
  & \multirow{2}{*}{Time (s)} \\
\cmidrule(lr){2-3}\cmidrule(lr){4-5}
  &
  Acc@5
  & $\overline{\mathrm{F1}}$
  & Acc@5
  & $\overline{\mathrm{F1}}$
  &
  & \\
\midrule

CoSIL
  & $79.2{\pm}0.3$
  & $29.8{\pm}0.1$
  & $24.6{\pm}0.4$
  & $12.9{\pm}0.2$
  & $24.3{\pm}0.0$
  & $33.3{\pm}0.1$ \\

NINJA (25k budget)
  & $74.2{\pm}0.5$
  & $48.4{\pm}0.4$
  & $63.4{\pm}0.6$
  & $22.5{\pm}0.3$
  & $24.8{\pm}0.1$
  & $35.6{\pm}0.4$ \\

OrcaLoca
  & $64.0{\pm}0.5$
  & $51.8{\pm}0.5$
  & $39.0{\pm}0.7$
  & $27.9{\pm}0.2$
  & $30.5{\pm}0.2$
  & $121.7{\pm}1.3$ \\

NINJA (30k budget)
  & $77.6{\pm}0.4$
  & $51.9{\pm}0.5$
  & $66.4{\pm}0.5$
  & $24.8{\pm}0.3$
  & $29.4{\pm}0.1$
  & $40.6{\pm}0.5$ \\

\bottomrule
\end{tabularx}

\end{table*}

\endgroup

\begingroup

\begin{table*}[t]
\small
\centering
\setlength{\tabcolsep}{3.5pt}
\renewcommand{\arraystretch}{1.15}

\newcolumntype{C}{>{\centering\arraybackslash}X}

\caption{Controlled comparison of framework designs and agentic fine-tuning strategies using Qwen3-Coder-30B-A3B-Instruct on SWE-Bench-Lite.}
\label{tab:finetuning_control}

\begin{tabularx}{\textwidth}{
  >{\raggedright\arraybackslash}p{1.8cm}
  >{\raggedright\arraybackslash}p{1.6cm}
  CCCCCC
}
\toprule
\multirow{2}{*}{Framework}
  & \multirow{2}{*}{Training}
  & \multicolumn{2}{c}{File-level}
  & \multicolumn{2}{c}{Function-level}
  & \multirow{2}{*}{\# Toks (k)}
  & \multirow{2}{*}{Time (s)} \\
\cmidrule(lr){3-4}\cmidrule(lr){5-6}
  &
  &
  Acc@5
  & $\overline{\mathrm{F1}}$
  & Acc@5
  & $\overline{\mathrm{F1}}$
  &
  & \\
\midrule

\multirow{3}{*}{OrcaLoca}
  & Initial
  & $64.0{\pm}0.5$
  & $51.8{\pm}0.5$
  & $39.0{\pm}0.7$
  & $27.9{\pm}0.2$
  & $30.5{\pm}0.2$
  & $121.7{\pm}1.3$ \\

  & SFT
  & $65.3{\pm}0.6$
  & $52.1{\pm}0.5$
  & $40.1{\pm}0.6$
  & $28.1{\pm}0.3$
  & $30.9{\pm}0.2$
  & $123.9{\pm}1.5$ \\

  & SFT+PPO
  & $74.5{\pm}0.5$
  & $54.9{\pm}0.4$
  & $45.5{\pm}0.5$
  & $28.4{\pm}0.3$
  & $28.8{\pm}0.2$
  & $116.9{\pm}1.4$ \\

\midrule

\multirow{5}{*}{NINJA}
  & Initial
  & $79.3{\pm}0.7$
  & $53.5{\pm}0.7$
  & $71.1{\pm}0.4$
  & $27.7{\pm}0.6$
  & $51.7{\pm}0.5$
  & $85.4{\pm}3.2$ \\

  & Direct PPO
  & $84.8{\pm}0.6$
  & $54.1{\pm}0.7$
  & $69.5{\pm}0.5$
  & $25.2{\pm}0.6$
  & $47.3{\pm}0.5$
  & $79.5{\pm}2.8$ \\

  & SFT
  & $82.3{\pm}0.5$
  & $53.8{\pm}0.6$
  & $71.8{\pm}0.4$
  & $27.6{\pm}0.5$
  & $51.0{\pm}0.5$
  & $96.8{\pm}3.4$ \\

  & SFT+DAPO
  & $88.2{\pm}0.5$
  & $58.7{\pm}0.6$
  & $74.5{\pm}0.5$
  & $28.3{\pm}0.6$
  & $47.7{\pm}0.5$
  & $80.5{\pm}2.7$ \\

\rowcolor[HTML]{EDEEEE}
  & SFT+PPO
  & $\mathbf{90.6{\pm}0.4}$
  & $\mathbf{59.7{\pm}0.5}$
  & $\mathbf{77.3{\pm}0.5}$
  & $\mathbf{29.4{\pm}0.6}$
  & $48.7{\pm}0.4$
  & $84.7{\pm}3.0$ \\

\bottomrule
\end{tabularx}
\vspace{-0.4cm}
\end{table*}

\endgroup

\textbf{Computational cost analysis.}
(1) As shown in Tab.~\ref{tab:ablation}, \textit{w/o dynamic scheduling} repeatedly dispatches inspectors to already explored files, increasing both time and token cost. Dynamic scheduling mitigates this redundancy by tracking explored files in the global state and removing them from subsequent dispatch candidates, thereby improving overall efficiency.
(2) As shown in Fig.~\ref{fig:parallelism_impact}, increasing inspector parallelism monotonically increases average runtime and token cost. This is because larger parallel batches reduce the frequency of global state updates, making it more likely that different inspectors explore overlapping files or functions and thus incur redundant computation. Meanwhile, a moderate amount of overlap can help inspectors recover potentially missed relevant functions, leading to improved performance as parallelism initially increases. However, when the parallelism degree becomes too high, excessive redundancy introduces noisy and repetitive evidence, which interferes with the agent's decision making and eventually degrades performance.


\subsection{Impact of token budget}

\textbf{Settings.}
To study the effect of token budget on our NINJA framework, we define the token budget $B$ as the maximum context length allowed for each independent agent context. Because the navigator and all inspectors maintain separate contexts, the same budget $B$ is applied independently to the navigator and to each inspector. Once an agent’s context exceeds $B$, it must terminate exploration early and return its result. 
Using Qwen3-Coder-30B-A3B-Instruct as the backbone, we evaluate NINJA on SWE-Bench-Lite under different token budgets $B$. We report function-level Acc@5, along with the token-budget exceed rates for navigator and inspector trajectories.

\begin{figure}[tbp]
  \centering
  \begin{subfigure}[b]{0.49\textwidth}
    \includegraphics[width=\textwidth]{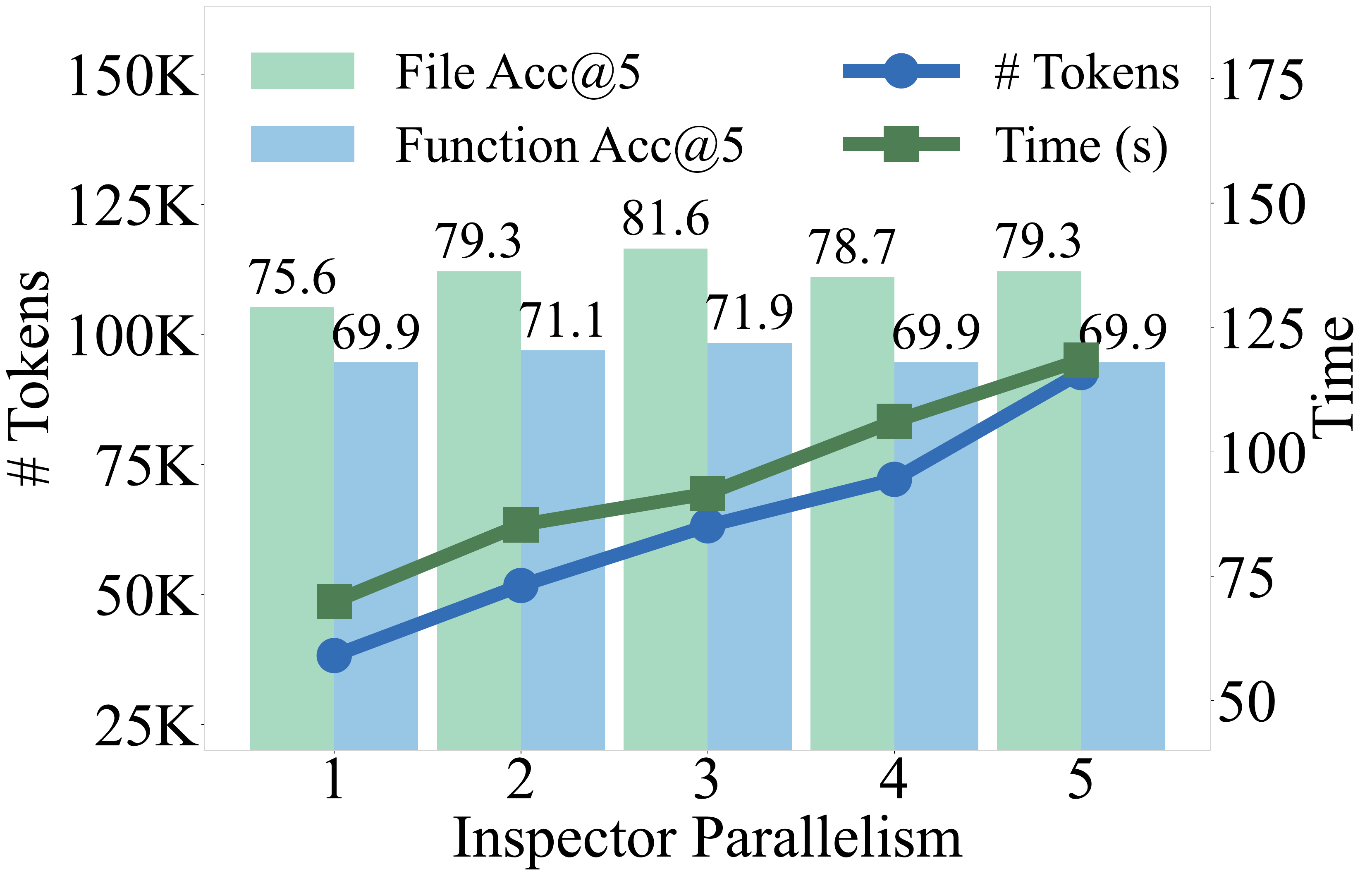}   
    \caption{Impact of inspector parallelism.}
    \label{fig:parallelism_impact}
  \end{subfigure}
  \hfill
  \begin{subfigure}[b]{0.49\textwidth}
    \includegraphics[width=\textwidth]{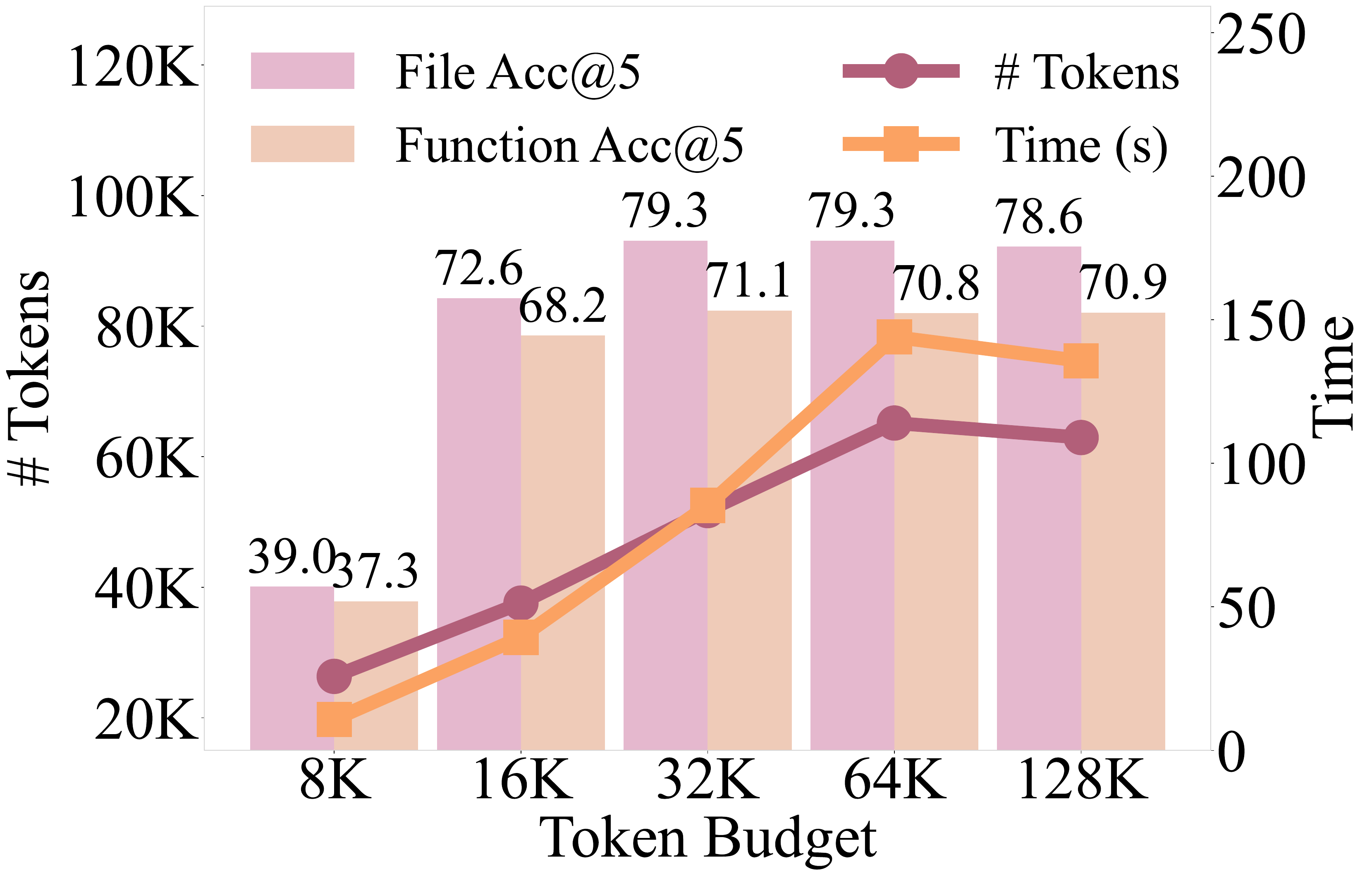}
    \caption{Impact of token budget.}
    \label{fig:token_budget_impact}
  \end{subfigure}
  \caption{Effects of inspector parallelism and token budget on NINJA's performance and efficiency.}
  
  \label{fig:parallelism_budget_impact}
  \vspace{-0.5cm}
\end{figure}

\textbf{Variation in performance with token budget.}
As shown in Fig.~\ref{fig:token_budget_impact}, the performance of NINJA improves as the token budget increases and gradually plateaus. To explain this trend, we further examine the token-budget exceed rates of the navigator and inspectors under different budgets. As shown in Tab.~\ref{tab:token_budget_exceed_rate} of Appendix, when the budget is small (<32K), both agents frequently exceed the limit, which forces early termination and leads to insufficient repository exploration. Once the budget reaches 32K or above, budget exceedance becomes rare, allowing the agents to complete most exploration trajectories. Consequently, NINJA's performance stabilizes and converges.

\subsection{Analysis of agentic fine-tuning}

\textbf{Settings.}
We analyze the proposed agentic fine-tuning pipeline by addressing four questions.
First, to study the effect of the critic model, we compare PPO with DAPO~\citep{yu2025dapo}, where PPO uses a critic model while DAPO follows a critic-free RL design.
Second, to examine the effect of SFT warmup, we compare direct RL training with RL training initialized from SFT-warmed models.
Third, to evaluate scalability across model sizes, we apply the same fine-tuning pipeline to both Qwen3-14B and Qwen3-Coder-30B-A3B-Instruct, and compare their post-training performance.
Finally, to assess the preservation of general reasoning ability, we evaluate the models before and after agentic fine-tuning on general reasoning benchmarks.

\textbf{Effect of the critic model.}
As shown in Tab.~\ref{tab:finetuning_control}, PPO outperforms DAPO on SWE-Bench-Lite, suggesting the importance of the critic model for long-horizon agentic RL.
Repository exploration trajectories are long and contain many intermediate decisions.
DAPO assigns the same trajectory-level advantage to all tokens, making it difficult to distinguish effective decisions from ineffective ones.
In contrast, PPO estimates token-level advantages with a critic model, providing finer-grained learning signals that help the model learn intermediate exploration decisions.

\textbf{Effect of SFT warmup.}
As shown in Tab.~\ref{tab:finetuning_control}, applying SFT warmup before RL yields stronger post-training performance. The initial model sometimes violates the required output format or produces tool calls that fail to execute. SFT warmup mitigates these issues by familiarizing the model with the tool-use protocol, providing a more stable policy initialization for subsequent RL.

\textbf{Generalization across model backbones.}
As shown in Tab.~\ref{tab:main_results_swe_bench_lite} and additional results in Appendix~\ref{app:locbench_results}, both NINJA-30B and NINJA-14B achieve clear performance gains after training.
This indicates that our agentic fine-tuning pipeline can generalize across different model scales.

\textbf{Preservation of general reasoning ability.}
As shown in Tab.~\ref{tab:general_reasoning_14b} for Qwen3-14B and Tab.~\ref{tab:general_reasoning_30b} for Qwen3-Coder-30B-A3B-Instruct, the trained models largely preserve their original performance on general reasoning benchmarks.
These results suggest that our task-specific agentic fine-tuning does not noticeably compromise the models' general reasoning capabilities.

\FloatBarrier
\Needspace{8\baselineskip}
\subsection{Downstream utility of edit and support location hints}
\label{sec:downstream_utility}

We examine whether NINJA's localization predictions improve downstream issue resolution and whether additional contextual locations offer further benefits. Existing localization benchmarks primarily target \textit{edit locations}, i.e., code units modified by the reference patch~\citep{chen2025locagent}. However, producing a correct patch may also require understanding \textit{support locations}: code units that are not edited but provide relevant context, such as related callee functions.

\begin{wraptable}{r}{0.65\textwidth}
\centering
\small
\setlength{\tabcolsep}{2pt}
\renewcommand{\arraystretch}{1.15}
\captionsetup{font=small,skip=4pt}
\caption{Downstream issue resolution on SWE-Bench-Verified with mini-SWE-agent and Claude-Haiku-4.5. Token usage is averaged per issue.}
\label{tab:downstream_resolution_hints}
\begin{tabularx}{\linewidth}{>{\raggedright\arraybackslash}Xlcc}
\toprule
\textbf{Hint source} & \textbf{Locations} & \textbf{Resolve (\%)} & \textbf{\# Toks (k)} \\
\midrule
None & --- & 28.8 & 39.3 \\
\midrule
\multirow{2}{*}{Ground truth} & Edit & 52.6 & 36.7 \\
 & Edit + support & 57.2 & 36.3 \\
\midrule
NINJA & Edit & 45.2 & 37.8 \\
\mbox{\tabindent w/o independent context} & Edit & 33.2 & 39.7 \\
\bottomrule
\end{tabularx}
\end{wraptable}

\textbf{Settings.}
We evaluate mini-SWE-agent~\citep{yang2024sweagent} with Claude-Haiku-4.5 on SWE-Bench-Verified~\citep{jimenez2023swe}. We compare five settings: no hints; predicted edit hints from NINJA; predicted edit hints from its \textit{w/o independent context} variant; ground-truth edit hints; and ground-truth edit+support hints. Predicted hints evaluate the practical utility of localized edit locations. Ground-truth hints isolate the contributions of edit locations and support locations to issue resolution. We extract support locations from reference patches following Appendix~\ref{app:support_location_extraction}. All hints are provided as file/function locations in the user prompt, as detailed in Appendix~\ref{app:prompts}. We report the resolve rate and average token usage per issue for the resolution agent.

\textbf{Predicted edit hints improve issue resolution.}
As shown in Tab.~\ref{tab:downstream_resolution_hints}, NINJA's predicted hints increase the resolve rate from 28.8\% to 45.2\%, a gain of 16.4 percentage points over no hints. They also outperform hints from the \textit{w/o independent context} variant (33.2\%) by 12.0 points. Meanwhile, NINJA's hints reduce the resolution agent's average token usage from 39.3k to 37.8k. These results show that NINJA's localization gains translate into more successful downstream repairs.

\textbf{Support locations provide complementary context.}
Ground-truth edit hints raise the resolve rate to 52.6\%, leaving a 7.4-point gap relative to NINJA's predicted hints. Adding ground-truth support locations further increases the resolve rate to 57.2\% (+4.6 points), while average token usage decreases from 36.7k to 36.3k.
These gains indicate that support locations provide complementary contextual information beyond edited locations, helping the issue resolution agent better understand related code dependencies and generate correct patches with higher effectiveness and efficiency. Therefore, current localization benchmarks should incorporate support locations into the ground-truth targets to better reflect the contextual requirements of real-world issue resolution.

\section{Conclusion}

In this work, we introduce NINJA, a hierarchical navigator--inspector joint architecture for issue localization. NINJA coordinates a global navigator with multiple dynamically scheduled inspectors for function-level local exploration, addressing two key limitations of existing agent-based methods: long-context accumulation and the difficulty of balancing repository-level breadth with function-level depth. We further apply a two-stage agentic fine-tuning strategy to strengthen both global coordination and local exploration. As a result, NINJA achieves highly competitive performance against strong baselines. Notably, after agentic fine-tuning, Qwen3-Coder-30B-A3B-Instruct surpasses the strong closed-source Claude-Haiku-4.5 model on both SWE-Bench-Lite and LocBench.

\section*{Acknowledgements}
This work is supported in part by the Hangzhou Joint Funds of the Zhejiang Provincial Natural Science Foundation of China under Grant No. LHZSD24F020001, in part by the Fundamental Research Funds for the Central Universities under Grant No. 226-2025-00057, and in part by the National Science Centre, Poland (NCN) project 2025/59/B/ST6/01950). This research is supported by the RIE2025 Industry Alignment Fund – Industry Collaboration Projects (IAF-ICP) (Award I2301E0026), administered by A*STAR, as well as by Alibaba Group and NTU Singapore through Alibaba-NTU Global e-Sustainability CorpLab (ANGEL).


{\small
\bibliographystyle{unsrtnat}
\bibliography{ref}
}
\newpage
\appendix
\part*{Appendix}
\vspace*{20pt}
\section*{Table of contents}
\hypersetup{
  linkcolor=darkblue
}
\startcontents[sections]
\printcontents[sections]{l}{1}{\setcounter{tocdepth}{2}}

\newpage
\section{Additional experimental results}
\label{app:additional_exp_results}
\subsection{Additional results on LocBench}
\label{app:locbench_results}

Beyond SWE-Bench-Lite, we further evaluate our method and the baselines on LocBench, as shown in Tab.~\ref{tab:main_results_locbench}. The results are consistent with those on SWE-Bench-Lite, confirming that NINJA maintains strong localization performance across different benchmarks.

\subsection{Further analysis of token budget scaling}
\label{app:exceed_rate_analysis}

To further analyze the token budget scaling behavior, we extract navigator and inspector trajectories from NINJA's inference runs on SWE-Bench-Lite. For each agent type, we count the number of trajectories that are forced to terminate early due to exceeding the token budget, and compute the corresponding exceed rate. As shown in Tab.~\ref{tab:token_budget_exceed_rate}, the exceed rates of both the navigator and the inspector decrease rapidly as the token budget increases. When the budget reaches 32K, budget exceedance becomes rare for both agents, suggesting that further increasing the token budget provides limited additional benefit.

\subsection{Effect of training on general reasoning ability}
\label{app:reasoning_generalization_analysis}
As shown in Tabs.~\ref{tab:general_reasoning_30b} and~\ref{tab:general_reasoning_14b}, models trained with our two-stage agentic fine-tuning largely preserve their original performance on math and general reasoning benchmarks (MATH500~\citep{lightman2023lets}, AIME24~\citep{aime24}, AIME25~\citep{aime25}, AMC23~\footnote{\url{https://huggingface.co/datasets/zwhe99/amc23}}, GPQA-Diamond~\citep{rein2024gpqa}, and MMLU-STEM, the STEM subset of MMLU~\citep{hendrycks2021mmlu}). This indicates that the proposed task-specific fine-tuning does not compromise the models' general reasoning ability while improving their issue localization performance. Notably, Qwen3-Coder-30B-A3B-Instruct even shows slight gains after training. One possible explanation is that the instruct-tuned model may underutilize its latent reasoning capability in the original format, whereas our think-then-act agentic training format encourages more explicit intermediate reasoning before tool use and decision making. These results suggest that our training strategy improves task-specific agentic behavior without causing evident degradation in general reasoning capacity.




\begingroup

\begin{table*}[thbp]
\small
\centering
\setlength{\tabcolsep}{6pt}
\renewcommand{\arraystretch}{1.15}

\newcolumntype{C}{>{\centering\arraybackslash}X}
\newcolumntype{P}[1]{>{\raggedright\arraybackslash}p{#1}}

\caption{Results on LocBench. Best and second-best localization results are in \textbf{bold} and \underline{underlined}, respectively. CodeRank, Haiku-4.5, and Qwen3-30B denote CodeRankEmbed, Claude-Haiku-4.5, and Qwen3-Coder-30B-A3B-Instruct. NINJA-14B and NINJA-30B are obtained by applying agentic fine-tuning (SFT+PPO) to Qwen3-14B and Qwen3-30B, respectively.}
\label{tab:main_results_locbench}

\begin{tabularx}{\textwidth}{P{1.6cm}P{1.55cm}CCCCCC}
\toprule
\multirow{2}{*}{Framework}
  & \multirow{2}{*}{Model}
  & \multicolumn{2}{c}{File-level}
  & \multicolumn{2}{c}{Function-level}
  & \multirow{2}{*}{\# Toks (k)}
  & \multirow{2}{*}{Time (s)} \\
\cmidrule(lr){3-4}\cmidrule(lr){5-6}
  &
  & Acc@5
  & F1
  & Acc@5
  & F1
  &
  & \\
\midrule

Embedding
  & CodeRank
  & $76.4{\pm}0.2$
  & $29.0{\pm}0.1$
  & $45.5{\pm}0.2$
  & $21.2{\pm}0.1$
  & N/A
  & $9.0{\pm}0.0$ \\

\midrule

\rowcolor[HTML]{EDEEEE}
  & Haiku-4.5
  & $40.7{\pm}0.3$
  & $18.7{\pm}0.2$
  & $21.3{\pm}0.4$
  & $10.9{\pm}0.2$
  & $55.6{\pm}0.2$
  & $157.7{\pm}0.6$ \\
\rowcolor[HTML]{EDEEEE}
\multirow{-2}{*}{SWERank+}
  & Qwen3-30B
  & $35.6{\pm}0.0$
  & $14.3{\pm}0.0$
  & $7.8{\pm}0.1$
  & $5.0{\pm}0.1$
  & $71.5{\pm}0.0$
  & $93.4{\pm}0.2$ \\

\multirow{2}{*}{LocAgent}
  & Haiku-4.5
  & $60.2{\pm}0.4$
  & $22.9{\pm}0.2$
  & $25.9{\pm}0.4$
  & $11.9{\pm}0.1$
  & $47.1{\pm}0.2$
  & $50.3{\pm}0.5$ \\
  & Qwen3-30B
  & $44.2{\pm}2.4$
  & $15.9{\pm}1.0$
  & $22.3{\pm}1.7$
  & $9.9{\pm}0.8$
  & $55.6{\pm}0.3$
  & $62.1{\pm}1.4$ \\

\rowcolor[HTML]{EDEEEE}
  & Haiku-4.5
  & $79.6{\pm}0.3$
  & $31.6{\pm}0.2$
  & $45.4{\pm}0.4$
  & $26.8{\pm}0.1$
  & $26.4{\pm}0.1$
  & $93.4{\pm}0.3$ \\
\rowcolor[HTML]{EDEEEE}
\multirow{-2}{*}{CoSIL}
  & Qwen3-30B
  & $79.0{\pm}0.3$
  & $30.7{\pm}0.0$
  & $19.1{\pm}0.9$
  & $15.8{\pm}0.2$
  & $27.1{\pm}0.1$
  & $30.5{\pm}0.5$ \\

\multirow{2}{*}{OrcaLoca}
  & Haiku-4.5
  & $59.1{\pm}0.1$
  & $59.2{\pm}0.1$
  & $36.6{\pm}0.3$
  & $32.0{\pm}0.2$
  & $33.2{\pm}0.3$
  & $153.1{\pm}3.4$ \\
  & Qwen3-30B
  & $50.7{\pm}0.4$
  & $51.3{\pm}0.4$
  & $24.8{\pm}0.3$
  & $23.6{\pm}0.3$
  & $26.6{\pm}0.2$
  & $51.7{\pm}3.1$ \\

\rowcolor[HTML]{EDEEEE}
  & Haiku-4.5
  & $81.5{\pm}0.4$
  & $31.4{\pm}0.1$
  & $22.1{\pm}0.4$
  & $15.2{\pm}0.2$
  & $26.8{\pm}0.1$
  & $115.1{\pm}0.4$ \\
\rowcolor[HTML]{EDEEEE}
\multirow{-2}{*}{RepoSearcher}
  & Qwen3-30B
  & $71.3{\pm}0.6$
  & $28.5{\pm}0.3$
  & $17.3{\pm}1.0$
  & $10.8{\pm}0.2$
  & $30.2{\pm}0.0$
  & $60.6{\pm}1.5$ \\

\midrule

\multirow{5}{*}{NINJA (Ours)}
  & Haiku-4.5
  & $\underline{83.3{\pm}0.4}$
  & $\mathbf{65.1{\pm}0.4}$
  & $\underline{65.8{\pm}0.5}$
  & $\underline{35.8{\pm}0.2}$
  & $44.7{\pm}0.2$
  & $148.3{\pm}0.6$ \\
  & Qwen3-14B
  & $71.3{\pm}0.2$
  & $53.3{\pm}0.1$
  & $46.9{\pm}0.4$
  & $26.7{\pm}0.1$
  & $40.1{\pm}0.5$
  & $55.5{\pm}0.6$ \\
  & NINJA-14B
  & $\underline{83.3{\pm}0.3}$
  & $55.7{\pm}0.3$
  & $50.8{\pm}0.4$
  & $27.3{\pm}0.2$
  & $35.0{\pm}0.2$
  & $45.7{\pm}0.3$ \\
  & Qwen3-30B
  & $74.5{\pm}0.7$
  & $58.4{\pm}0.7$
  & $60.3{\pm}0.4$
  & $33.6{\pm}0.6$
  & $59.8{\pm}0.5$
  & $89.4{\pm}3.2$ \\
  & NINJA-30B
  & $\mathbf{86.7{\pm}0.3}$
  & $\underline{63.7{\pm}0.3}$
  & $\mathbf{69.0{\pm}0.4}$
  & $\mathbf{37.5{\pm}0.2}$
  & $57.9{\pm}0.2$
  & $88.7{\pm}0.5$ \\

\bottomrule
\end{tabularx}

\end{table*}

\endgroup
\begin{table}[t] 
\centering 
\small 
\caption{Token budget exceed rates of navigator and inspector.} \label{tab:token_budget_exceed_rate} 
\begin{tabular}{lccccc} 
\toprule 
Agent & 8K & 16K & 32K & 64K & 128K \\ 
\midrule 
Navigator & 82.19\% & 13.04\% & 0.33\% & 0.00\% & 0.00\% \\ 
Inspector & 49.83\% & 12.90\% & 0.46\% & 0.00\% & 0.00\% \\ 
\bottomrule 
\end{tabular} 
\end{table}

\begin{table*}[t]
\centering
\small
\setlength{\tabcolsep}{6pt}
\renewcommand{\arraystretch}{1.15}
\caption{General reasoning performance of Qwen3-Coder-30B-A3B-Instruct before and after agentic fine-tuning. Scores are averaged over four sampling runs (Avg@4). GPQA* denotes GPQA-Diamond, and MMLU* denotes MMLU-STEM.}
\label{tab:general_reasoning_30b}
\begin{tabular}{llccccccc}
\toprule
Agent & Settings 
& MATH500 
& AIME24 
& AIME25 
& AMC23 
& GPQA*
& MMLU*
& Avg. \\
\midrule
& initial & 88.9 & 28.3 & 20.8 & 68.8 & 49.9 & 91.0 & 58.0 \\
\midrule
\multirow{2}{*}{Navigator}
& SFT & 89.0 & 33.3 & 28.3 & 73.1 & 48.6 & 91.0 & 60.6 \\
& SFT+PPO & 89.0 & 35.0 & 25.8 & 75.0 & 48.9 & 90.9 & 60.8 \\
\midrule
\multirow{2}{*}{Inspector}
& SFT & 89.1 & 31.7 & 29.2 & 71.3 & 48.9 & 90.9 & 60.2 \\
& SFT+PPO & 89.4 & 30.9 & 28.4 & 71.9 & 48.4 & 90.9 & 60.0 \\
\bottomrule
\end{tabular}
\end{table*}
\begin{table*}[t]
\centering
\small
\setlength{\tabcolsep}{6pt}
\renewcommand{\arraystretch}{1.15}
\caption{General reasoning performance of Qwen3-14B before and after agentic fine-tuning. Scores are averaged over four sampling runs (Avg@4). GPQA* denotes GPQA-Diamond, and MMLU* denotes MMLU-STEM.}
\label{tab:general_reasoning_14b}
\begin{tabular}{llccccccc}
\toprule
Agent & Settings 
& MATH500 
& AIME24 
& AIME25 
& AMC23 
& GPQA*
& MMLU*
& Avg. \\
\midrule
& Initial & 97.3 & 80.8 & 75.0 & 98.8 & 65.8 & 94.1 & 85.3 \\
\midrule
\multirow{2}{*}{Navigator}
& SFT & 97.3 & 81.7 & 73.3 & 98.1 & 65.8 & 94.1 & 85.1 \\
& SFT+PPO & 97.0 & 80.8 & 73.7 & 98.8 & 66.3 & 94.0 & 85.1 \\
\midrule
\multirow{2}{*}{Inspector}
& SFT & 97.7 & 82.5 & 72.5 & 99.4 & 66.5 & 94.1 & 85.5 \\
& SFT+PPO & 97.0 & 79.7 & 71.7 & 100.0 & 65.0 & 93.9 & 84.6 \\
\bottomrule
\end{tabular}
\end{table*}

\section{Implementation details}
\label{sec:implementation_details}
\subsection{Details of support location extraction}
\label{app:support_location_extraction}

As defined in Sec.~\ref{sec:backgrounds}, a location is represented as $(f_i, u_j)$, where $f_i$ denotes a file and $u_j$ denotes a code unit. A statement line refers to a concrete line of code inside a code unit $u_j$. In practice, resolving an issue often requires reasoning beyond the functions that are directly edited by the patch. For example, suppose \texttt{func1} calls \texttt{func2} at a statement line inside \texttt{func1}, and the patch modifies this call site. Since the edited statement belongs to \texttt{func1}, \texttt{func1} is included in the edited-location set. However, because no line inside \texttt{func2} is directly modified, \texttt{func2} is not included as an edited location. Nevertheless, understanding the behavior or constraints of \texttt{func2} may be essential for producing the correct patch. We therefore define such repository locations as \textit{support locations}: they are not directly edited by the patch, but provide necessary contextual information for issue resolution.

We extract support locations from patches based on repository-defined code units that are directly referenced by edited statements within the edit locations. Specifically, if an edited statement invokes a repository-defined function or method, we include the invoked code unit as a support location. This covers regular function calls, class-method calls, class instantiations, and object-call expressions. Class instantiations are mapped to their constructors, while object-call expressions are mapped to the corresponding \texttt{\_\_call\_\_} method when resolvable.

For example, if a patch adds, removes, or modifies a call such as \texttt{validate\_config(config)}, we include \texttt{validate\_config}. If it edits \texttt{response.raise\_for\_status()}, we include \texttt{Response.raise\_for\_status} when the receiver type can be resolved. If it edits \texttt{client = HttpClient(base\_url, timeout=30)}, we include \texttt{HttpClient.\_\_init\_\_}. Similarly, if it edits \texttt{result = model(x)}, we include \texttt{Model.\_\_call\_\_} when \texttt{model} can be resolved to a repository-defined class. In all cases, the support location must be resolvable within the target repository.

\subsection{Training settings}
\label{app:train_settings}
We implement both SFT and RL training using the verl framework~\citep{sheng2024hybridflow}.
\subsubsection{SFT warmup}

To enable Qwen3-14B and Qwen3-Coder-30B-A3B-Instruct to reliably follow the required output format, we construct a small SFT dataset from SWE-smith-py. Specifically, we sample 500 instances from SWE-smith-py and use Claude-Haiku-4.5 to generate agent trajectories. We then filter the trajectories based on task success and strict adherence to the required output format, yielding 462 valid navigator trajectories and 275 valid inspector trajectories. The SFT hyperparameters are reported in Tab.~\ref{tab:sft_hyperparams}. To mitigate overfitting given the limited size of the filtered dataset, we train each model for two epochs. All SFT experiments are conducted on 8 NVIDIA A800 GPUs.

\subsubsection{Reinforcement learning}
For RL, we sample at most 300 instances from each repository in SWE-smith-py, yielding 24,209 training instances in total. This sampling strategy exposes the agents to diverse repositories and encourages them to learn search strategies that generalize across different codebases. We compare two RL algorithms, PPO and DAPO: PPO uses a critic model for value estimation, whereas DAPO is critic-free. The training hyperparameters for PPO and DAPO are reported in Tab.~\ref{tab:ppo_hyperparams} and Tab.~\ref{tab:dapo_hyperparameters}, respectively. For DAPO, we set the training batch size to 64 and generate 4 samples per instance, making its effective rollout batch size comparable to PPO, which uses a training batch size of 256 with 1 sample per instance. 
We train Qwen3-14B on 8 NVIDIA H100 GPUs and train Qwen3-Coder-30B-A3B-Instruct using distributed multi-node training across two H100 machines.
\begin{table}[t]
\centering
\small
\caption{Training hyperparameters of SFT.}
\label{tab:sft_hyperparams}
\begin{tabular}{ll}
\toprule
\textbf{Hyperparameter} & \textbf{Value} \\
\midrule
Global batch size & 32 \\
Micro batch size per GPU & 2 \\
Training epochs & 2 \\
Maximum sequence length & 32,768 \\
Maximum tokens per GPU & 32,768 \\
Learning rate & $5\times 10^{-6}$ \\
Learning rate scheduler & Cosine \\
Warmup ratio & 0.1 \\
Minimum learning rate ratio & 0.1 \\
Weight decay & 0.01 \\
Gradient clipping & 1.0 \\
\bottomrule
\end{tabular}
\end{table}
\begin{table}[t]
\centering
\small
\caption{Training hyperparameters of PPO.}
\label{tab:ppo_hyperparams}
\begin{tabular}{ll}
\toprule
\textbf{Hyperparameter} & \textbf{Value} \\
\midrule
Training batch size & 256 \\
PPO mini-batch size & 64 \\
Actor learning rate & $1\times 10^{-6}$ \\
Critic learning rate & $1\times 10^{-5}$ \\
Critic warmup steps & 50 \\
Discount factor $\gamma$ & 1.0 \\
GAE parameter $\lambda$ & 1.0 \\
KL loss coefficient & 0.001 \\
Value clipping range & 0.5 \\
Sampling temperature & 0.6 \\
Sampling top-$p$ & $0.95$ \\
Samples per prompt & 1 \\
Maximum prompt length & 8,000 \\
Maximum response length & 24,768 \\
\bottomrule
\end{tabular}
\end{table}
\begin{table*}[t]
\centering
\small
\caption{Training hyperparameters of DAPO.}
\label{tab:dapo_hyperparameters}
\begin{tabular}{ll}
\toprule
\textbf{Hyperparameter} & \textbf{Value} \\
\midrule
Training batch size & $64$  \\
PPO mini-batch size & $64$ \\
KL loss coefficient &  $0.001$ \\
Clip ratio & $[0.20, 0.28]$ \\
Samples per prompt & $4$ \\
Sampling temperature & $0.6$ \\
Sampling top-$p$ & $0.95$ \\
Learning rate & $1\times 10^{-6}$ \\
Warmup steps & $10$ \\
Maximum prompt length & $8{,}000$ \\
Maximum response length & $24{,}768$ \\
\bottomrule
\end{tabular}
\end{table*}

\subsection{Agent tools}
\label{app:tools}
\begin{table}[t]
\centering
\caption{Repository tools for the navigator agent.}
\label{tab:navigator_tools}
\small
\setlength{\tabcolsep}{6pt}
\renewcommand{\arraystretch}{1.15}

\begin{tabularx}{\linewidth}{@{}L{3.4cm}Y@{}}
\toprule
\textbf{Tool} & \textbf{Description} \\
\midrule

\texttt{find\_files\_by\_content}
& Given keywords or regex patterns and an optional file pattern, searches repository contents and returns matched files with relevant line snippets.
\fulldashline

\texttt{find\_files\_by\_name}
& Given filename or path patterns, matches repository files and returns a deduplicated list of candidate file paths.
\fulldashline

\texttt{dispatch}
& Given selected entry files, dispatches inspector agents for in-depth exploration and returns suspicious locations with supporting exploration evidence.
\fulldashline

\texttt{exit}
& Signals that the navigator has completed repository search and is ready to produce the final localization result. \\

\bottomrule
\end{tabularx}
\end{table}
\begin{table}[t]
\centering
\caption{Repository tools for the inspector agent.}
\label{tab:inspector_tools}
\small
\setlength{\tabcolsep}{6pt}
\renewcommand{\arraystretch}{1.15}

\begin{tabularx}{\linewidth}{@{}L{4.0cm}Y@{}}
\toprule
\textbf{Tool} & \textbf{Description} \\
\midrule

\texttt{get\_methods\_of\_class}
& Given a file and class name, returns all method signatures defined in the class, helping the inspector identify candidate methods for further inspection.
\fulldashline

\texttt{get\_file\_functions}
& Given a file name, returns all module-level function signatures in the file, providing a quick overview of its functional structure.
\fulldashline

\texttt{get\_file\_classes}
& Given a file name, returns all class signatures in the file, helping the inspector locate potentially relevant classes.
\fulldashline

\texttt{get\_code\_of\_function}
& Given a file and function name, returns the source code of the specified module-level function for detailed analysis.
\fulldashline

\texttt{get\_code\_of\_class\_method}
& Given a file, class name, and method name, returns the source code of the specified class method for detailed analysis.
\fulldashline

\texttt{get\_callers}
& Given a function or method location, returns caller locations, enabling upstream usage tracing.
\fulldashline

\texttt{get\_callees}
& Given a function or method location, returns callee locations, enabling downstream dependency tracing.
\fulldashline

\texttt{exit}
& Signals that exploration is complete and returns suspicious locations together with supporting evidence. \\

\bottomrule
\end{tabularx}
\end{table}

As shown in Tab.~\ref{tab:navigator_tools}, the navigator is equipped with file-level search tools and a dispatch tool, which allow it to identify candidate files and assign them as entry files to inspectors. As shown in Tab.~\ref{tab:inspector_tools}, the inspector is equipped with function-level inspection tools and dependency-tracing tools, enabling it to examine issue-relevant functions along caller--callee chains.

\subsection{Details of evaluation settings}
\label{app:evaluation_metrics}

For each benchmark, we independently run the evaluation three times and report the mean and standard deviation.

\textbf{Evaluation hyperparameters.}
For inference, we set the temperature to $0.6$ and top-$p$ to $0.95$. For Claude-Haiku-4.5, top-$p$ is left unspecified. Unless otherwise stated, the context budget for each independent context is $32{,}768$ tokens, and inspector parallelism is set to $P=2$.

\textbf{Metrics.}
For performance, we report Acc@$k$ and $\overline{\mathrm{F1}}$. Let the evaluation set contain $I$ instances. For the $i$-th instance, let $\hat{\mathcal{G}}_i$ denote the ranked list of predicted issue-relevant locations and $\mathcal{G}^{\star}_i$ denote the corresponding ground-truth locations, where each location follows the function-level representation defined above. We compute function-level metrics directly on these locations. For file-level metrics, we first project both predicted and ground-truth locations to their file components and then apply the same evaluation procedure.

For Acc@$k$, we keep only the top-$k$ predictions:
$
\hat{\mathcal{G}}_i^{(k)}
=
\hat{\mathcal{G}}_i[:k].
$
An instance is counted as correct if all ground-truth locations are covered by the top-$k$ predictions:
\begin{equation}
\mathrm{Acc@}k
=
\frac{1}{I}
\sum_{i=1}^{I}
\mathbf{1}
\left[
\mathcal{G}^{\star}_i \subseteq \hat{\mathcal{G}}_i^{(k)}
\right],
\end{equation}
where $\mathbf{1}[\cdot]$ is the indicator function. We report both file-level and function-level Acc@5.

For F1, we compute the instance-level set F1 defined in Eq.~\eqref{eq:f1} and then average it over all instances:
\begin{equation}
\overline{\mathrm{F1}}
=
\frac{1}{I}
\sum_{i=1}^{I}
\mathrm{F1}(\hat{\mathcal{G}}_i, \mathcal{G}^{\star}_i).
\end{equation}
For efficiency, we report runtime and token cost. Runtime is measured as the end-to-end inference time of the localization process, including model calls and intermediate processing, while excluding offline pre-processing such as embedding construction, retrieval-index tokenization, or graph-index construction. 
Token cost is measured as the total trajectory length of all LLM calls involved in localization, including both navigator and inspector trajectories for NINJA.

\subsection{Prompts}
\label{app:prompts}
The system and user prompts for the navigator and inspector agents are presented in Boxes~B.5.1--B.5.4. Box~B.5.5 shows the location-hinted user prompt template used for mini-SWE-agent.

\begin{tcolorbox}[
    float*,
    floatplacement=t,
    colback=blue!5!white,
    colframe=blue!75!black,
    title=Box B.5.1: System prompt of navigator,
    fonttitle=\bfseries,
    boxrule=0.8pt,
    arc=2mm,
    width=\textwidth,
    label={box:navigator_system_prompt}
]
\small
\ttfamily
\obeylines
You are the navigator agent for issue localization that coordinates the search for suspicious code locations by dispatching inspector agents to explore promising files in a code repository.
~
\# Tools
You may call one or more functions to assist with the user query.
You are provided with function signatures within <tools></tools> XML tags:
~
\textcolor{red}{\textbf{\{\{Tool Schemas\}\}}}
\end{tcolorbox}

\begin{tcolorbox}[
    float*,
    floatplacement=t,
    colback=teal!5!white,
    colframe=teal!55!black,
    title=Box B.5.2: System prompt of inspector,
    fonttitle=\bfseries,
    boxrule=0.7pt,
    arc=2mm,
    width=\textwidth,
    label={box:inspector_system_prompt}
]
\small
\ttfamily
\obeylines
You are the inspector agent for issue localization that explores a specific file entry point in a code repository to identify suspicious functions/methods related to a reported issue.
~
\# Tools
You may call one or more functions to assist with the user query.
You are provided with function signatures within <tools></tools> XML tags:
~
\textcolor{red}{\textbf{\{\{Tool Schemas\}\}}}
\end{tcolorbox}
\begin{tcolorbox}[
    float*,
    floatplacement=t,
    colback=blue!5!white,
    colframe=blue!75!black,
    title=Box B.5.3: User prompt of navigator,
    fonttitle=\bfseries,
    boxrule=0.8pt,
    arc=2mm,
    width=\textwidth,
    breakable,  
    label={box:navigator_user_prompt}
]

\small
\ttfamily
\obeylines
\textless{}pr\_description\textgreater{}
Consider the following PR description:
\textcolor{red}{\$problem\_statement}
\textless{}/pr\_description\textgreater{}
~
\textless{}repo\_structure\textgreater{}
The repository directory structure is as follows:
\textcolor{red}{\$repo\_structure}
\textless{}/repo\_structure\textgreater{}
~
\textless{}instructions\textgreater{}
\# Task Instructions
\#\# Overview
You are coordinating a multi-agent search to localize the root cause of a reported issue in a code repository.
Your goal is to locate all functions/methods that are related to and potentially causing the reported issue.
Use `find\_files\_by\_content' and `find\_files\_by\_name' to identify suspicious files as entry points, then use `dispatch' to send them to inspector agents for in-depth exploration.
Integrate feedback from inspector agents and decide whether to expand the search or terminate.
~
For each response:
1. Include a \textless{}think\textgreater{} section explaining your reasoning and what you are trying to accomplish.
2. And then you can provide one or more \textless{}tool\_call\textgreater{} blocks to execute.
~
\#\# Recommended Workflow
1. Analyze the issue description to extract keywords and understand the bug symptoms.
2. Use `find\_files\_by\_content' and `find\_files\_by\_name' to search for relevant files based on extracted keywords and the repository structure.
3. Once you have found one or more plausible entry files, use `dispatch' to send them to inspector agents.
4. Read and integrate inspector agent results. Decide whether the suspicious locations sufficiently cover all issue symptoms.
5. If important symptoms remain uncovered, search for new keywords or bridging modules and dispatch additional entry files.
6. When suspicious locations are sufficient or no more productive exploration is possible, call `exit` to finish.
~
\#\# Current State
After each tool execution, the system will report the current state with the following variables:
- `confirmed\_suspicious': Functions/methods identified so far as potentially causing the issue. These are automatically accumulated from inspector agent results.
- `explored\_entries': Files that have already been explored by inspector agents. These will be skipped if dispatched again.
- `pending\_entries': Files discovered by `find\_files\_by\_content` / `find\_files\_by\_name` that are waiting for you to dispatch to inspector agents via `dispatch`. You should dispatch these or search for more files.
~
\textless{}IMPORTANT\textgreater{}
- In each response, perform only the next best action. Do not narrate the full workflow or mention future steps.
- Prefer action over explanation.
- Focus on finding the ROOT CAUSE, not just symptoms.
- Stop exploring when you have found relevant suspicious locations.
\textless{}/IMPORTANT\textgreater{}
~
\#\# Allowed Response Formats
You must choose exactly one format in each response.
~
\#\#\# Format A: Continue searching
Use this format only when you still need more information or need to dispatch inspector agents.
~
\textless{}think\textgreater{}
...
\textless{}/think\textgreater{}
\textless{}tool\_call\textgreater{}
\{"name": "...", "arguments": \{...\}\}
\textless{}/tool\_call\textgreater{}
~
You may include multiple \textless{}tool\_call\textgreater{} blocks.
~
\#\#\# Format B: Finish the search
Use this format only when you are ready to stop searching.
~
\textless{}think\textgreater{}
...
\textless{}/think\textgreater{}
\textless{}tool\_call\textgreater{}
\{"name": "exit", "arguments": \{\}\}
\textless{}/tool\_call\textgreater{}
~
Do not include any non-exit tool call in this format.
~
\#\# Examples of CORRECT Responses
\textless{}example\_response\textgreater{}
\textless{}think\textgreater{}
The issue mentions a "timeout" error in HTTP requests. I should search for files related to HTTP client and timeout configuration.
\textless{}/think\textgreater{}
\textless{}tool\_call\textgreater{}
\{"name": "find\_files\_by\_content", "arguments": \{"keywords": ["timeout"], "file\_pattern": "*.py"\}\}
\textless{}/tool\_call\textgreater{}
\textless{}/example\_response\textgreater{}
~
\textless{}example\_response\textgreater{}
\textless{}think\textgreater{}
I found several promising files related to the HTTP client. I should dispatch them to inspector agents for in-depth exploration.
\textless{}/think\textgreater{}
\textless{}tool\_call\textgreater{}
\{"name": "dispatch", "arguments": \{"entries": ["src/client/http.py", "src/config/timeout.py"]\}\}
\textless{}/tool\_call\textgreater{}
\textless{}/example\_response\textgreater{}
~
\textless{}example\_response\textgreater{}
\textless{}think\textgreater{}
The inspector agents have found sufficient suspicious locations that cover all issue symptoms. I can now finish the search.
\textless{}/think\textgreater{}
\textless{}tool\_call\textgreater{}
\{"name": "exit", "arguments": \{\}\}
\textless{}/tool\_call\textgreater{}
\textless{}/example\_response\textgreater{}
~
\textless{}/instructions\textgreater{}
\end{tcolorbox}
\lstset{
    basicstyle=\ttfamily\small,  
    breaklines=true,              
    breakatwhitespace=false,      
    columns=flexible,
    keepspaces=true,
    aboveskip=0pt,   
    belowskip=0pt,   
}

\begin{tcolorbox}[
    float*,
    floatplacement=t,
    colback=teal!5!white,
    colframe=teal!55!black,
    title=Box B.5.4: User prompt of inspector,
    fonttitle=\bfseries,
    boxrule=0.7pt,
    arc=2mm,
    width=\textwidth,
    breakable,
    label={box:inspector_user_prompt}
]
\small
\ttfamily
\obeylines
\textless{}pr\_description\textgreater{}
Consider the following PR description:
\textcolor{red}{\$problem\_statement}
\textless{}/pr\_description\textgreater{}
~
\textless{}entry\_file\textgreater{}
Start your investigation from the following entry file:
\textcolor{red}{\$entry\_file}
\textless{}/entry\_file\textgreater{}
~
\textless{}instructions\textgreater{}
\# Task Instructions
~
\#\# Overview
You are exploring a specific file entry point to find functions/methods that are suspicious and potentially related to the reported issue.
Use tools to inspect code, trace call relationships, and identify the root cause.
You are NOT limited to the entry file — follow promising leads across files.
Final locations must be at function-level granularity: a module-level function or a class method.
~
For each response:
1. Include a `\textless{}think\textgreater{}' section explaining your reasoning and what you are trying to accomplish.
2. If you still need more information, provide one or more `\textless{}tool\_call\textgreater{}' blocks to execute.
3. If you are ready to finish, provide a `\textless{}result\textgreater{}' block and also include an `exit()' tool call in the same response.
~
\#\# Recommended Workflow
1. Start by examining the entry file: use `get\_file\_functions' and `get\_file\_classes' to understand its structure.
2. Based on the issue description, identify which functions/classes look suspicious.
3. Use `get\_code\_of\_function' or `get\_code\_of\_class\_method' to inspect suspicious code in detail.
4. Use `get\_callers' and `get\_callees' to trace call relationships and discover related code.
5. Follow promising leads across files until you find the root cause.
6. When confident, return the final localization result. If nothing is suspicious, return an empty result.
~
\textless{}IMPORTANT\textgreater{}
- Focus on functions/methods that are MOST related to the issue symptoms.
- When you find suspicious code, explain WHY it is suspicious in relation to the issue.
- Do not explore endlessly — be efficient and focused.
- You MUST return at least 5 suspicious locations. Explore broadly enough to find multiple related functions/methods.
- The final response must include both the `\textless{}result\textgreater{}' block and the `exit()' tool call.
- If no relevant location exists, return an empty `suspicious' list.
\textless{}/IMPORTANT\textgreater{}
~
\#\# Allowed Response Formats
You must choose exactly one format in each response.
~
\#\#\# Format A: Continue searching
Use this format only when you still need more information.
~
\textless{}think\textgreater{}
...
\textless{}/think\textgreater{}
\textless{}tool\_call\textgreater{}
\{"name": "...", "arguments": \{...\}\}
\textless{}/tool\_call\textgreater{}
~
You may include multiple \textless{}tool\_call\textgreater{} blocks.
Do not include \textless{}result\textgreater{} in this format.
~
\#\#\# Format B: Finish exploration
Use this format only when you are ready to stop exploring.
You MUST include at least 5 suspicious locations in the result.
~
\textless{}think\textgreater{}
...
\textless{}/think\textgreater{}
\textless{}result\textgreater{}
\begin{lstlisting}
{
  "suspicious": [
    {"location": "file_path:function_name", "reason": "explanation of why this is suspicious"},
    {"location": "file_path:ClassName.method_name", "reason": "explanation of why this is suspicious"},
    {"location": "file_path:function_name2", "reason": "explanation"},
    {"location": "file_path:ClassName2.method_name2", "reason": "explanation"},
    {"location": "file_path:function_name3", "reason": "explanation"}
  ]
}
\end{lstlisting}
\textless{}/result\textgreater{}
\textless{}tool\_call\textgreater{}
\{"name": "exit", "arguments": \{\}\}
\textless{}/tool\_call\textgreater{}
~
Do not include any non-exit tool call in this format.
~
The `location' format must be:
- `file\_path:function\_name' for module-level functions
- `file\_path:ClassName.method\_name' for class methods
~
\#\# Examples of CORRECT Responses
\textless{}example\_response\textgreater{}
\textless{}think\textgreater{}
I should start by understanding the structure of the entry file to identify potentially suspicious functions.
\textless{}/think\textgreater{}
\textless{}tool\_call\textgreater{}
\{"name": "get\_file\_functions", "arguments": \{"file\_name": "src/client/http.py"\}\}
\textless{}/tool\_call\textgreater{}
\textless{}tool\_call\textgreater{}
\{"name": "get\_file\_classes", "arguments": \{"file\_name": "src/client/http.py"\}\}
\textless{}/tool\_call\textgreater{}
\textless{}/example\_response\textgreater{}
~
\textless{}example\_response\textgreater{}
\textless{}think\textgreater{}
The `send\_request' method handles timeout logic. Based on the issue, the timeout value is not being passed correctly, which matches the reported bug symptoms. I traced the call chain and found multiple related functions.
\textless{}/think\textgreater{}
\textless{}result\textgreater{}
\begin{lstlisting}
{
  "suspicious": [
    {"location": "src/client/http.py:HttpClient.send_request", "reason": "Timeout parameter is ignored when retry is enabled, causing the default timeout to be used instead"},
    {"location": "src/config/settings.py:load_timeout_config", "reason": "Returns None instead of the default timeout value when config key is missing"},
    {"location": "src/client/http.py:HttpClient._prepare_request", "reason": "Does not propagate timeout from session config to the underlying transport"},
    {"location": "src/client/retry.py:RetryPolicy.execute", "reason": "Overrides user-specified timeout with a hardcoded default on each retry attempt"},
    {"location": "src/client/http.py:HttpClient.__init__", "reason": "Default timeout attribute is set to None instead of reading from config"}
  ]
}
\end{lstlisting}
\textless{}/result\textgreater{}
\textless{}tool\_call\textgreater{}
\{"name": "exit", "arguments": \{\}\}
\textless{}/tool\_call\textgreater{}
\textless{}/example\_response\textgreater{}
~
\textless{}/instructions\textgreater{}
\end{tcolorbox}
\begin{tcolorbox}[
    float*,
    floatplacement=t,
    colback=orange!6!white,
    colframe=orange!75!black,
    title=Box B.5.5: User prompt of mini-SWE-Agent,
    fonttitle=\bfseries,
    boxrule=0.8pt,
    arc=2mm,
    width=\textwidth,
    breakable,  
    label={box:miniswe_user_prompt}
]
\small
\ttfamily
\obeylines
    <pr\_description>
    Consider the following PR description:
    
    \textcolor{red}{\{\{task\}\}}
    </pr\_description>
    ~
    \{\% if edit\_hint\_locations or support\_hint\_locations \%\}
    <hint\_locations>
    The following file/function locations were identified through prior issue localization.
    \{\% if edit\_hint\_locations \%\}
    <edit\_locations>
    Code units directly modified by the fix — your primary edit targets:
    \textcolor{red}{\{\{edit\_hint\_locations\}\}}
    </edit\_locations>
    \{\% endif \%\}
    \{\% if support\_hint\_locations \%\}
    <support\_locations>
    Code units NOT directly edited but required for understanding the issue (related functions or methods):
    \textcolor{red}{\{\{support\_hint\_locations\}\}}
    </support\_locations>
    \{\% endif \%\}
    </hint\_locations>
    \{\% endif \%\}
~
    <instructions>
    \# Task Instructions
~
    \#\# Overview
    You're a software engineer interacting continuously with a computer by submitting commands.
    You'll be helping implement necessary changes to meet requirements in the PR description.
    Your task is specifically to make changes to non-test files in the current directory in order to fix the issue described in the PR description in a way that is general and consistent with the codebase.
    ~
    <IMPORTANT>
    This is an interactive process where you will think and issue AT LEAST ONE command, see the result, then think and issue your next command(s). 
    </IMPORTANT>
~
    For each response:
    1. Include a THOUGHT section explaining your reasoning and what you're trying to accomplish
    2. Provide one or more bash tool calls to execute
~
    \#\# Hint Usage
    \{\% if edit\_hint\_locations or support\_hint\_locations \%\}
    You are given location hints from prior issue localization. There are two types:
    \{\% if edit\_hint\_locations \%\}
    - **Edit locations**: code units directly modified by the fix. Start your investigation here — these are the most likely places to make changes.
    \{\% endif \%\}
    \{\% if support\_hint\_locations \%\}
    - **Support locations**: code units that are NOT directly edited but provide essential context (related functions or methods). Read these to understand how the edit locations are used before making changes.
    \{\% endif \%\}
~
    Use these hints early when deciding what files to inspect first.
    Treat them as strong search priors, not as guaranteed ground truth.
    Verify each hinted location by reading the repository before making edits.
    If the hints conflict with the code or the PR description, trust the code and the PR description.
    \{\% else \%\}
    No extra location hints are available for this task.
    \{\% endif \%\}
~
    \#\# Important Boundaries
    - MODIFY: Regular source code files in /testbed (this is the working directory for all your subsequent commands)
    - DO NOT MODIFY: Tests, configuration files (pyproject.toml, setup.cfg, etc.)
~
    \#\# Recommended Workflow
    1. Analyze the codebase by finding and reading relevant files
    2. Create a script to reproduce the issue
    3. Edit the source code to resolve the issue
    4. Verify your fix works by running your script again
    5. Test edge cases to ensure your fix is robust
~
    \#\# Command Execution Rules
    You are operating in an environment where
    1. You issue at least one command
    2. The system executes the command(s) in a subshell
    3. You see the result(s)
    4. You write your next command(s)
~
    Each response should include:
    1. **Reasoning text** where you explain your analysis and plan
    2. At least one tool call with your command
~
    **CRITICAL REQUIREMENTS:**
    - Your response SHOULD include reasoning text explaining what you're doing
    - Your response MUST include AT LEAST ONE bash tool call. You can make MULTIPLE tool calls in a single response when the commands are independent (e.g., searching multiple files, reading different parts of the codebase).
    - Directory or environment variable changes are not persistent. Every action is executed in a new subshell.
    - However, you can prefix any action with `MY\_ENV\_VAR=MY\_VALUE cd /path/to/working/dir \&\& ...' or write/load environment variables from files
~
    Example of a CORRECT response:
    <example\_response>
    I need to understand the Builder-related code. Let me find relevant files and check the project structure.
~
    [Makes multiple bash tool calls: \{"command": "ls -la"\}, \{"command": "find src -name '*.java' | grep -i builder"\}, \{"command": "cat README.md | head -50"\}]
    </example\_response>
~
    \#\# Environment Details
    - You have a full Linux shell environment
    - Always use non-interactive flags (-y, -f) for commands
    - Avoid interactive tools like vi, nano, or any that require user input
    - You can use bash commands or invoke any tool that is available in the environment
    - You can also create new tools or scripts to help you with the task
    - If a tool isn't available, you can also install it
~
    \#\# Submission
    When you've completed your work, you MUST submit your changes as a git patch.
    Follow these steps IN ORDER, with SEPARATE commands:
~
    Step 1: Create the patch file
    Run `git diff -- path/to/file1 path/to/file2 > patch.txt' listing only the source files you modified.
    Do NOT commit your changes.
    <IMPORTANT>
    The patch must only contain changes to the specific source files you modified to fix the issue.
    Do not submit file creations or changes to any of the following files:
    - test and reproduction files
    - helper scripts, tests, or tools that you created
    - installation, build, packaging, configuration, or setup scripts unless they are directly part of the issue you were fixing (you can assume that the environment is already set up for your client)
    - binary or compiled files
    </IMPORTANT>
~
    Step 2: Verify your patch
    Inspect patch.txt to confirm it only contains your intended changes and headers show `--- a/' and `+++ b/' paths.
~
    Step 3: Submit (EXACT command required)
    You MUST use this EXACT command to submit:
    ```bash
    echo COMPLETE\_TASK\_AND\_SUBMIT\_FINAL\_OUTPUT \&\& cat patch.txt
    ```
    If the command fails (nonzero exit status), it will not submit.
~
    <CRITICAL>
    - Creating/viewing the patch and submitting it MUST be separate commands (not combined with \&\&).
    - If you modify patch.txt after verifying, you SHOULD verify again before submitting.
    - You CANNOT continue working (reading, editing, testing) in any way on this task after submitting.
    </CRITICAL>
    ~
    </instructions>
\end{tcolorbox}


\section{Implementation details of baselines}
\label{app:baselines_comparison}
\textbf{CodeRankEmbed.} CodeRankEmbed uses a bi-encoder initialized from Arctic-Embed-M and contrastively fine-tuned on the CoRNStack dataset~\citep{suresh2024cornstack} using InfoNCE loss. Given an issue description, it encodes both the query and repository-level code entities into a shared embedding space and returns the top-ranked candidates by cosine similarity.

\textbf{SWERank+.} SWERank+~\citep{reddy2025swerank+} combines cross-lingual code ranking with a multi-turn agent loop. It retrieves candidate locations using an embedding-based retriever and a listwise LLM reranker, while the agent iteratively rewrites search queries based on previous results and maintains a persistent memory buffer to refine the localization output.

\textbf{LocAgent.} LocAgent~\citep{chen2025locagent} is a graph-guided LLM agent for issue localization. It parses repositories into directed heterogeneous graphs whose nodes represent files, classes, and functions, and whose edges encode dependency relations such as imports and invocations. These graphs enable agent to perform multi-hop reasoning, therefore identifying the root cause of issue.

\textbf{CoSIL.} CoSIL~\citep{jiang2025cosil} employs a two-phase code graph search strategy. First, it reduces the search space by exploring file-level module call graphs. Second, it iteratively searches over function call graphs to collect relevant contexts. The LLM dynamically constructs call graphs during search, while a context-pruning mechanism controls the search direction and manages context length.

\textbf{OrcaLoca.} OrcaLoca~\citep{yu2025orcaloca} first indexes the repository into a code graph, then lets the agent explore it through a reason-and-act loop. To decide what to search next, it uses a priority queue that ranks the candidate actions by relevance. When a search returns a large class or file, it automatically breaks it down into individual methods and scores their relevance before further exploration. To keep the context clean, it prunes away search results that are far from the suspected bug locations on the code graph.

\textbf{RepoSearcher.} RepoSearcher~\citep{ma2025tool} is a single-agent localization method fine-tuned in two stages. It first performs rejection-sampled SFT on the model's own successful trajectories, followed by RL using nDCG@$k$ as the reward. During localization, the agent uses five tools to retrieve repository structure, file imports, and code entity definitions.

All baselines are open-sourced on GitHub. We use their released implementations and follow the original parameter settings to evaluate Qwen3-Coder-30B-A3B-Instruct and Claude-Haiku-4.5 on SWE-Bench-Lite and LocBench. For a fair comparison, we use the same context token budget of $32{,}768$, temperature of $0.6$, and top-$p$ of $0.95$ during evaluation.
We do not retrain LocAgent or RepoSearcher. LocAgent trains its localization model by distilling trajectories sampled from Claude-3.5-Sonnet. Since the distilled model is not expected to surpass its Claude teacher, further distilling Claude-Haiku-4.5 trajectories into a Qwen-based LocAgent would at best approximate Claude-Haiku-4.5 performance. RepoSearcher relies on a 28K training dataset that has not been publicly released, so we do not reproduce its training stage. Instead, we evaluate RepoSearcher using its lightweight toolset framework as a baseline.

\section{Additional related work}
\label{app:additional_related_work}
\subsection{Multi-agent frameworks for software engineering}
\label{app:se_multiagent_frameworks}
Multi-agent frameworks have shown strong potential in software engineering by decomposing complex development and maintenance workflows across specialized roles. Representative systems such as ChatDev~\citep{qian2024chatdev} and MetaGPT~\citep{hong2024metagpt} target end-to-end software development, coordinating role-specialized agents across stages such as requirements and design, coding, testing, and documentation. AutoCodeRover~\citep{zhang2024autocoderover}, MAGIS~\citep{tao2024magis}, and CodeR~\citep{chen2024coder} instead focus on repository-level issue resolution.
In contrast to both lines of work, NINJA specifically targets issue localization, a prerequisite for downstream issue resolution, rather than the complete software development or repair workflow. Moreover, their localization mechanisms differ substantially from NINJA.
AutoCodeRover performs structure-aware code search over AST-based program representations and can optionally incorporate spectrum-based fault localization; MAGIS ranks files using BM25, filters them through LLM-based relevance assessment, and reuses file summaries through repository-evolution memory; CodeR combines BM25 relevance with spectrum-based suspiciousness scores derived from reproduction tests. Although MAGIS and CodeR use multiple agents across the overall issue-resolution workflow, each assigns its dedicated localization stage to a single specialized role---the Repository Custodian and Fault Localizer, respectively. In contrast, NINJA organizes localization itself as multi-round collaboration between two specialized agent roles: the navigator maintains global file-level search state and dynamically dispatches multiple inspectors, while each inspector performs function-level exploration in an independent context and can trace dependencies across files.

\section{Discussion}
\label{app:discussion}
\textbf{Locating edited and support locations.}
The results in Sec.~\ref{sec:downstream_utility} provide initial evidence that support locations can offer complementary contextual information beyond edited locations and benefit downstream issue resolution. However, consistent with existing issue localization benchmarks, our training and evaluation primarily use edited locations as localization targets. As a result, the localization of support locations remains largely unexplored. Compared with function-level edited locations, support locations require more fine-grained reasoning over the concrete edited statements. Specifically, a method must first identify the statements likely to be modified, determine whether they invoke or depend on other repository functions, and then retrieve the corresponding support locations. This makes support-location localization more challenging because it requires statement-level evidence tracking in addition to function-level localization. Designing effective and efficient methods for jointly locating edited and support locations remains an important direction for future work.

\textbf{Cost--performance trade-off.}
NINJA incurs additional localization cost, but the downstream results in Tab.~\ref{tab:downstream_resolution_hints} indicate that this cost yields practically meaningful gains. Predicted edit-location hints from the lower-cost \textit{NINJA w/o independent context} variant improve the resolve rate only from 28.8\% without hints to 33.2\%. In contrast, NINJA's predicted hints raise the resolve rate to 45.2\%, approaching the 52.6\% obtained with ground-truth edit-location hints. These results indicate that less accurate localization can directly constrain downstream issue resolution. We therefore view NINJA as a higher-effort localization option, analogous to configurable higher-effort modes in coding agents such as Codex~\citep{openai2026codex} and Claude Code~\citep{claude2025claudecode}: users can accept additional computation time in exchange for better coding outcomes when resolution quality is the priority, while selecting lower-cost alternatives when latency or cost constraints dominate.

\textbf{Limitations.}
NINJA has two main limitations. First, its hierarchical multi-agent search introduces additional runtime overhead compared with simpler baselines, although the total time cost remains within roughly twice that of other agent-based methods. Second, our experiments are conducted on Python repositories. Nevertheless, the tool design of NINJA is not Python-specific: the navigator--inspector architecture and the separation between file-level search and function-level inspection can be extended to repositories in other programming languages, provided that language-specific parsing and dependency-inspection tools are available.

\section{Broader impact}
\label{app:broader_impact}
This work studies issue localization, whose core capability is to explore a repository and identify code locations relevant to a natural-language issue description. Beyond issue resolution, this capability can support a range of repository-level software engineering tasks. For repository-level question answering, it can retrieve concrete functions, classes, and files as evidence for answering code-related questions, rather than relying only on surface-level semantic similarity. For code understanding, it can help developers trace cross-file dependencies, caller--callee relations, and execution-relevant code paths in large codebases. For repository documentation generation, it can identify the code units that implement a feature or behavior and organize them into structured explanations. More broadly, accurate issue localization can reduce the manual effort required to navigate unfamiliar repositories and can make downstream repair, testing, and documentation agents more efficient by narrowing the search space. Since repository exploration may also expose sensitive implementation details or security-relevant code paths, such systems should be applied only to authorized repositories, and their predictions should be treated as decision support rather than fully autonomous conclusions.

\section{Case study}
\label{app:case_study}
\definecolor{thinkcolor}{RGB}{230, 244, 230}
\definecolor{toolcolor}{RGB}{204, 229, 255}
\definecolor{answergreen}{RGB}{0, 150, 0}
\definecolor{answerred}{RGB}{200, 0, 0}
\definecolor{headercolor}{RGB}{255, 230, 200}
\definecolor{obscolor}{RGB}{255, 235, 205}
\definecolor{answercolor}{RGB}{255, 220, 150}
\definecolor{thinkbg}{RGB}{220, 240, 220}
\definecolor{toolbg}{RGB}{220, 235, 255}
\definecolor{issuebg}{RGB}{255, 245, 230}
\definecolor{answerbg}{RGB}{210, 235, 210}
\definecolor{passgreen}{RGB}{0, 140, 0}
\definecolor{failred}{RGB}{180, 0, 0}
\definecolor{highlightblue}{RGB}{0, 80, 180}
\definecolor{highlightred}{RGB}{180, 0, 0}

\newtcolorbox{sysbox}{
  colback=answercolor,
  colframe=answercolor,
  arc=2pt,
  boxrule=0pt,
  left=4pt, right=4pt, top=3pt, bottom=3pt,
  fontupper=\small,
  before upper={\raggedright}
}

\newtcolorbox{thinkbox}{
  colback=thinkcolor,
  colframe=thinkcolor,
  arc=2pt,
  boxrule=0pt,
  left=4pt, right=4pt, top=3pt, bottom=3pt,
  fontupper=\small,
  before upper={\raggedright}
}

\newtcolorbox{toolbox}{
  colback=toolcolor,
  colframe=toolcolor,
  arc=2pt,
  boxrule=0pt,
  left=4pt, right=4pt, top=3pt, bottom=3pt,
  fontupper=\small,
  before upper={\raggedright}
}

\newtcolorbox{obsbox}{
  colback=toolcolor,
  colframe=toolcolor,
  arc=2pt,
  boxrule=0pt,
  left=4pt, right=4pt, top=3pt, bottom=3pt,
  fontupper=\small,
  before upper={\raggedright}
}

\newtcolorbox{NavigatorBox}[1][]{
  colback=issuebg, colframe=orange!70!red, boxrule=0.8pt,
  arc=3pt, left=6pt, right=6pt, bottom=4pt,
  fontupper=\small,
  before upper={\raggedright},
  coltitle=black,
  fonttitle=\bfseries,
  halign title=center,
  #1
}

\tcbset{
  casebox/.style={
    arc=2pt, boxrule=0pt,
    left=4pt, right=4pt, top=2pt, bottom=2pt,
    fontupper=\small, before upper={\setlength{\parskip}{2pt}},
  }
}

\begin{figure*}[tp]
\centering
 
\begin{NavigatorBox}[title=Navigator]
\begin{minipage}[t]{0.48\linewidth}
\textbf{Repo:} django/django \\
\textbf{Instance ID:} django\_\_django-11019 \\
\textbf{search\_execution\_mode:} multi\_agent \\[2pt]
\textbf{<SYSTEM>}\\
You are the navigator agent for issue localization that coordinates the search for suspicious code locations by dispatching inspector agents to explore promising files in a code repository.\\ [2pt]
\# Tool Definition: \\
-\texttt{find\_files\_by\_content} \\
-\texttt{find\_files\_by\_name} \\
-\texttt{dispatch} \\
-\texttt{exit} \\ [2pt]
\textbf{<USER>} \\ [2pt]
\# Issue Description:\\
Merging 3 or more media objects can throw unnecessary \texttt{MediaOrderConflictWarnings}.
Consider the following form definition, where text-editor-extras.js depends on text-editor.js but all other JS files are independent: \\[2pt]
\# Repository Structure \\
...

\end{minipage}
\hfill\vrule\hfill
\begin{minipage}[t]{0.48\linewidth}
\textbf{<think>}\\
Merging 3+ media objects throws unnecessary \texttt{MediaOrderConflictWarning}. The JS file ordering is incorrectly flagged as conflicting despite being valid. Need to locate where the warning is raised — starting with \texttt{forms/widgets.py} — and trace through the media merge logic.\\ [2pt]
\textbf{<tool\_call>} \\ 
\path{find_files_by_content(}\\
\path{"MediaOrderConflictWarning",}\\
\path{"media.*order.*conflict")} \\[1pt]
\path{find_files_by_name(}\\
\path{"**/forms/widgets.py")}\\ [2pt]
\textbf{<observation>} \\
matched files: \\
1. \path{django/contrib/gis/forms/widgets.py} \\
2. \path{django/forms/widgets.py} \\ [2pt]
\textbf{<think>} \\
Now I'll dispatch the main \texttt{forms/widgets.py} file to a inspector agent... \\ [2pt]
\textbf{<tool\_call>} \\
\path{dispatch(}\\
\path{"django/contrib/gis/forms/widgets.py",}\\
\path{"django/forms/widgets.py")}
\end{minipage}
\end{NavigatorBox}
\vspace{10pt}
 
\begin{minipage}[t]{0.48\textwidth}
  \begin{tcolorbox}[casebox, colback=white, colframe=gray!40, boxrule=0.5pt]
    \centering\small\textbf{Inspector\_0}
  \end{tcolorbox}
  \begin{sysbox}
      \textbf{<SYSTEM>} \\
      You are an inspector agent for issue localization that explores a specific file entry point in a code repository to identify suspicious functions/methods related to a reported issue.\\
      \textbf{<issue\_description>} \\
      ...\\
      \textbf{<entry\_file>}\\
      \path{django/forms/widgets.py} \\
      ...
  \end{sysbox}
  \begin{thinkbox}
      \textbf{<think>}\\
      Starting from the entry file to understand its structure and identify potentially suspicious functions related to the media ordering issue.
 
      \textbf{<tool\_call>}\\
      \path{get_file_classes(django/forms/widgets.py)}\\
      \path{get_file_functions(django/forms/widgets.py)}

      \textbf{<observation>} \\
      Module-level functions: \\
      \texttt{media\_property, boolean\_check}. \\
      Classes include: \\
      \texttt{MediaOrderConflictWarning, Media, MediaDefiningClass}.
  \end{thinkbox}

\end{minipage}
\hfill
\begin{minipage}[t]{0.48\textwidth}
 
  \begin{tcolorbox}[casebox, colback=white, colframe=gray!40, boxrule=0.5pt]
    \centering\small\textbf{Inspector\_1}
  \end{tcolorbox}
 
  \begin{sysbox}
      \textbf{<SYSTEM>} \\
      You are an inspector agents for issue localization that explores a specific file entry point in a code repository to identify suspicious functions/methods related to a reported issue.\\
      \textbf{<issue\_description>} \\
      ...\\
      \textbf{<entry\_file>}\\
      \path{django/contrib/gis/forms/widgets.py} \\
      ...
  \end{sysbox}
  \begin{thinkbox}
      \textbf{<think>}\\
      Need to examine the entry file's structure to identify potentially suspicious functions related to the media merging issue.
 
      \textbf{<tool\_call>}\\
      \path{get_file_functions(django/contrib/gis/forms/widgets.py)} \\
      \path{get_file_classes(django/contrib/gis/forms/widgets.py)}

      \textbf{<observation>} \\
      No module-level functions.  \\
      Three widget classes found: \\
      \texttt{BaseGeometryWidget, OpenLayersWidget, OSMWidget}. Need to inspect their methods for media-related logic.
  \end{thinkbox}

\end{minipage}


\caption{Case study of issue \textit{django\_\_django-11019} from SWE-Bench-Lite. (I)}
\label{fig:case_study_0}
\end{figure*}

\definecolor{thinkcolor}{RGB}{230, 244, 230}
\definecolor{toolcolor}{RGB}{204, 229, 255}
\definecolor{obscolor}{RGB}{255, 235, 205}
\definecolor{answercolor}{RGB}{255, 220, 150}
\definecolor{issuebg}{RGB}{255, 245, 230}
\definecolor{answerbg}{RGB}{210, 235, 210}
\definecolor{thinkbg}{RGB}{220, 240, 220}
\definecolor{toolbg}{RGB}{220, 235, 255}
\definecolor{passgreen}{RGB}{0, 140, 0}
\definecolor{failred}{RGB}{180, 0, 0}
\definecolor{answergreen}{RGB}{0, 150, 0}
\definecolor{answerred}{RGB}{200, 0, 0}
\definecolor{headercolor}{RGB}{255, 230, 200}
\definecolor{highlightblue}{RGB}{0, 80, 180}
\definecolor{highlightred}{RGB}{180, 0, 0}






\tcbset{
  casebox/.style={
    arc=2pt, boxrule=0pt,
    left=4pt, right=4pt, top=2pt, bottom=2pt,
    fontupper=\small, before upper={\setlength{\parskip}{2pt}},
  }
}

\begin{figure*}[tp]
\centering

\begin{minipage}[t]{0.48\textwidth}
  \begin{tcolorbox}[casebox, colback=white, colframe=gray!40, boxrule=0.5pt]
    \centering\small\textbf{Inspector\_0}
  \end{tcolorbox}
  \begin{toolbox}
  \textbf{<think>}\\
    The \texttt{Media} class is the primary suspect. Inspecting its methods to identify merge-related logic.
  
      \textbf{<tool\_call>}\\
      \texttt{get\_methods\_of\_class(Media)}

      \textbf{<observation>} \\
      Methods include \texttt{merge}, \texttt{\_\_add\_\_}, \texttt{\_js}, \texttt{\_css} — all involved in media combination. \\
      \texttt{merge} and \texttt{\_\_add\_\_} are the primary candidates.
  \end{toolbox}
  \begin{thinkbox}
      \textbf{<think>}\\
      \texttt{\_\_add\_\_} is likely involved in merging media objects and could be where conflict detection is triggered. Inspecting its implementation first.
 
      \textbf{<tool\_call>}\\
      \path{get_code_of_class_method(Media.__add__)}

      \textbf{<observation>} \\
      \texttt{\_\_add\_\_} simply concatenates \texttt{\_css\_lists} and \texttt{\_js\_lists} with no ordering validation. The actual conflict detection must happen elsewhere — likely in merge.
  \end{thinkbox}

  \begin{toolbox}
      \textbf{<think>}\\
      Attempting to inspect \path{MediaOrderConflictWarning.__init__} to understand the warning definition — but the call returns an error. Pivoting to \texttt{media\_property} as the upstream entry point instead.
 
      \textbf{<tool\_call>} \\
      \path{get_code_of_class_method(}\\
      \path{MediaOrderConflictWarning.__init__)}

      \textbf{<observation>} \\
      \texttt{media\_property} calls \texttt{\_\_add\_\_} via
      \path{m + Media(definition)}, indirectly triggering merge. Confirmed as upstream entry of the merge pipeline.
  \end{toolbox}
    \begin{tcolorbox}[
      colback=white, colframe=white,
      arc=0pt, boxrule=0pt,
      left=0pt, right=0pt, top=5pt, bottom=5pt,
    ]
    \centering\ldots
    \end{tcolorbox}
 

 
\end{minipage}
\hfill
\begin{minipage}[t]{0.48\textwidth}
 
  \begin{tcolorbox}[casebox, colback=white, colframe=gray!40, boxrule=0.5pt]
    \centering\small\textbf{Inspector\_1}
  \end{tcolorbox}
 
 

  \begin{toolbox}
      Need to examine the widget classes to look for Media-related functionality since the issue is about media merging and ordering conflicts.
 
      \textbf{<tool\_call>}\\
      \path{get_methods_of_class(BaseGeometryWidget)}\\
      \path{get_methods_of_class(OpenLayersWidget)}\\
      \path{get_methods_of_class(OSMWidget)}

      \textbf{<observation>} \\
      All three classes only expose \texttt{\_\_init\_\_}, \\
      serialize, deserialize, \texttt{get\_context} — no media logic present. Must cross to: \\
      \texttt{django/forms/widgets.py}.
  \end{toolbox}
  \begin{thinkbox}
      \textbf{<think>}\\
      GIS widgets contain no media logic. The issue is about \texttt{Django} forms media merging, so the core logic must be in \texttt{django/forms/widgets.py}. Switching there to locate the Media class.
 
      \textbf{<tool\_call>}\\
      \path{get_file_classes(django/forms/widgets.py)}

      \textbf{<observation>} \\
      Confirmed \path{MediaOrderConflictWarning},
      \texttt{Media}, and \texttt{MediaDefiningClass} in this file. \texttt{Media} is the primary suspect — proceeding to inspect its methods.
  \end{thinkbox}
  \begin{toolbox}
      \textbf{<think>}\\
      Merge is confirmed as the bug site. Need to identify its direct callers to fully establish the trigger path.
 
      \textbf{<tool\_call>}\\
        \path{get_callers(django/forms/widgets.py:Media.merge)}

      \textbf{<observation>} \\
      Returns \texttt{Media.\_css} and \texttt{Media.\_js} as the sole callers of merge. Full chain confirmed: \\
      \texttt{media\_property} \(\rightarrow\) \texttt{\_\_add\_\_} \(\rightarrow\) \texttt{\_js, \_css} \(\rightarrow\) \texttt{merge} \(\rightarrow\) \texttt{warning}. \\
      Ready to conclude.
  \end{toolbox}
      \begin{tcolorbox}[
      colback=white, colframe=white,
      arc=0pt, boxrule=0pt,
      left=0pt, right=0pt, top=5pt, bottom=5pt,
    ]
    \centering\ldots
    \end{tcolorbox}
 

 
\end{minipage}

\caption{Case study of issue \textit{django\_\_django-11019} from SWE-Bench-Lite. (II)}
\label{fig:case_study_1}
\end{figure*}

\begin{figure*}[tp]
\centering

\begin{minipage}[t]{0.48\textwidth}
  \begin{tcolorbox}[casebox, colback=white, colframe=gray!40, boxrule=0.5pt]
    \centering\small\textbf{Inspector\_0}
  \end{tcolorbox}
  \begin{tcolorbox}[
    colback=white, colframe=white,
    arc=0pt, boxrule=0pt,
    left=0pt, right=0pt, top=5pt, bottom=5pt,
    ]
    \centering\ldots
  \end{tcolorbox}
  \begin{thinkbox}
      \textbf{<think>}\\
      The call chain is fully established: \\
      \texttt{MediaDefiningClass.\_\_new\_\_} \(\rightarrow\)  \\
      \texttt{media\_property} \(\rightarrow\) \texttt{Media.\_\_add\_\_} \(\rightarrow\) \\
      \texttt{Media.merge} \(\rightarrow\) \texttt{warning}. \\
      Sufficient evidence to conclude exploration.
 
      \textbf{<tool\_call>}\\
      \texttt{exit()} 

      \textbf{<observation>} \\
      Returned 5 suspicious locations:\\
        1. \texttt{Media.merge} \\
        2. \texttt{Media.\_\_add\_\_} \\
        3. \texttt{MediaOrderConflictWarning} \\
        4. \texttt{media\_property} \\
        5. \texttt{MediaDefiningClass.\_\_new\_\_} 
  \end{thinkbox}
 
\end{minipage}
\hfill
\begin{minipage}[t]{0.48\textwidth}
 
  \begin{tcolorbox}[casebox, colback=white, colframe=gray!40, boxrule=0.5pt]
    \centering\small\textbf{Inspector\_1}
  \end{tcolorbox}
  \begin{tcolorbox}[
    colback=white, colframe=white,
    arc=0pt, boxrule=0pt,
    left=0pt, right=0pt, top=5pt, bottom=5pt,
    ]
    \centering\ldots
  \end{tcolorbox}
  \begin{thinkbox}
      \textbf{<think>}\\
      All key locations identified across the complete call chain. Sufficient evidence to conclude exploration.
 
      \textbf{<tool\_call>}\\
      \texttt{exit()} 

      \textbf{<observation>} \\
      Returned 5 suspicious locations:\\
        1. \texttt{Media.merge} \\
        2. \texttt{Media.\_js} \\
        3. \texttt{Media.\_css}  \\
        4. \texttt{Media.\_\_add\_\_}  \\
        5. \texttt{media\_property}
  \end{thinkbox}
 
\end{minipage}

\vspace{4pt}

\begin{NavigatorBox}[title=Navigator]
    \textbf{<think>} \\
    Finish the search now. Based on all confirmed suspicious locations and the issue description, rank and provide the top-5 most likely culprit locations for the bug.

    \textbf{<trace\_locs>}\\
        1. \texttt{django/forms/widgets.py — function: Media.merge} $\checkmark$ \\
        2. \texttt{django/forms/widgets.py — function: Media.\_\_add\_\_ }  \\
        3. \texttt{django/forms/widgets.py — function: Media.\_js} $\checkmark$ \\
        4. \texttt{django/forms/widgets.py — function: Media.\_css} $\checkmark$ \\
        5. \texttt{django/forms/widgets.py — function: media\_property} 
\end{NavigatorBox}
\caption{Case study of issue \textit{django\_\_django-11019} from SWE-Bench-Lite. (III)}
\label{fig:case_study_2}
\end{figure*}

To demonstrate how NINJA performs issue localization, we present a case study on instance \textit{django\_\_django-11019} from SWE-Bench-Lite, using Qwen3-Coder-30B-A3B-Instruct as the backbone model. The case illustrates the trajectories of the navigator and its dispatched inspectors, as shown in Figs.~\ref{fig:case_study_0}--\ref{fig:case_study_2}. Rather than following a predefined workflow, the navigator and inspectors dynamically reason over the current search state and adaptively determine subsequent actions.
This case concerns a false positive \texttt{MediaOrderConflictWarning} triggered when merging three or more Django form widget media objects. The issue stems from an intermediate merge step that imposes a spurious ordering constraint.

The navigator begins by calling \texttt{find\_files\_by\_content} with media-order conflict keywords and further uses \texttt{find\_files\_by\_name} to locate form widget files, identifying two candidate files: \texttt{django/forms/widgets.py} and \texttt{django/contrib/gis/forms/widgets.py}. It then dispatches both files to inspectors, each operating in an independent context.

Inspector\_0 is assigned to \texttt{django/forms/widgets.py} and immediately locates the key media-related classes. It progressively narrows its focus from the \texttt{Media} class to the \texttt{\_\_add\_\_} and \texttt{merge} methods. The inspector finds that \texttt{\_\_add\_\_} performs plain list concatenation without ordering validation, indicating that the conflict detection logic lies in \texttt{merge}. It further traces the upstream invocation chain through \texttt{media\_property} and \texttt{MediaDefiningClass.\_\_new\_\_}, establishing the full invocation chain.

Inspector\_1 starts from \texttt{django/contrib/gis/forms/widgets.py} and determines that the GIS-specific widget classes contain no media-merging logic, prompting cross-file traversal to \texttt{django/forms/widgets.py}. It independently identifies \texttt{Media.merge} as the key suspicious location and further calls \texttt{get\_callers} on \texttt{Media.merge}, confirming that \texttt{Media.\_js} and \texttt{Media.\_css} are its sole direct callers. This caller tracing allows NINJA to identify all three ground-truth locations.

The navigator aggregates the results from both inspectors and updates the global search state. Based on this state, it determines that the collected evidence is sufficient to cover the issue, ranks the confirmed suspicious locations, and outputs the following top-5 locations: \texttt{Media.merge}, \texttt{Media.\_\_add\_\_}, \texttt{Media.\_js}, \texttt{Media.\_css}, and \texttt{media\_property}. All three ground-truth locations---\texttt{Media.merge}, \texttt{Media.\_js}, and \texttt{Media.\_css}---are fully covered.


\newpage
\FloatBarrier
\section*{NeurIPS Paper Checklist}

\begin{enumerate}

\item {\bf Claims}
    \item[] Question: Do the main claims made in the abstract and introduction accurately reflect the paper's contributions and scope?
    \item[] Answer: \answerYes{} 
    \item[] Justification: The main claims clearly delineate the contributions and scope of this work.
    \item[] Guidelines:
    \begin{itemize}
        \item The answer \answerNA{} means that the abstract and introduction do not include the claims made in the paper.
        \item The abstract and/or introduction should clearly state the claims made, including the contributions made in the paper and important assumptions and limitations. A \answerNo{} or \answerNA{} answer to this question will not be perceived well by the reviewers. 
        \item The claims made should match theoretical and experimental results, and reflect how much the results can be expected to generalize to other settings. 
        \item It is fine to include aspirational goals as motivation as long as it is clear that these goals are not attained by the paper. 
    \end{itemize}

\item {\bf Limitations}
    \item[] Question: Does the paper discuss the limitations of the work performed by the authors?
    \item[] Answer: \answerYes{} 
    \item[] Justification: Appendix~\ref{app:discussion} discusses the limitations of this work.
    \item[] Guidelines:
    \begin{itemize}
        \item The answer \answerNA{} means that the paper has no limitation while the answer \answerNo{} means that the paper has limitations, but those are not discussed in the paper. 
        \item The authors are encouraged to create a separate ``Limitations'' section in their paper.
        \item The paper should point out any strong assumptions and how robust the results are to violations of these assumptions (e.g., independence assumptions, noiseless settings, model well-specification, asymptotic approximations only holding locally). The authors should reflect on how these assumptions might be violated in practice and what the implications would be.
        \item The authors should reflect on the scope of the claims made, e.g., if the approach was only tested on a few datasets or with a few runs. In general, empirical results often depend on implicit assumptions, which should be articulated.
        \item The authors should reflect on the factors that influence the performance of the approach. For example, a facial recognition algorithm may perform poorly when image resolution is low or images are taken in low lighting. Or a speech-to-text system might not be used reliably to provide closed captions for online lectures because it fails to handle technical jargon.
        \item The authors should discuss the computational efficiency of the proposed algorithms and how they scale with dataset size.
        \item If applicable, the authors should discuss possible limitations of their approach to address problems of privacy and fairness.
        \item While the authors might fear that complete honesty about limitations might be used by reviewers as grounds for rejection, a worse outcome might be that reviewers discover limitations that aren't acknowledged in the paper. The authors should use their best judgment and recognize that individual actions in favor of transparency play an important role in developing norms that preserve the integrity of the community. Reviewers will be specifically instructed to not penalize honesty concerning limitations.
    \end{itemize}

\item {\bf Theory assumptions and proofs}
    \item[] Question: For each theoretical result, does the paper provide the full set of assumptions and a complete (and correct) proof?
    \item[] Answer: \answerNA{} 
    \item[] Justification: This work does not include theory assumptions or proofs.
    \item[] Guidelines:
    \begin{itemize}
        \item The answer \answerNA{} means that the paper does not include theoretical results. 
        \item All the theorems, formulas, and proofs in the paper should be numbered and cross-referenced.
        \item All assumptions should be clearly stated or referenced in the statement of any theorems.
        \item The proofs can either appear in the main paper or the supplemental material, but if they appear in the supplemental material, the authors are encouraged to provide a short proof sketch to provide intuition. 
        \item Inversely, any informal proof provided in the core of the paper should be complemented by formal proofs provided in appendix or supplemental material.
        \item Theorems and Lemmas that the proof relies upon should be properly referenced. 
    \end{itemize}

    \item {\bf Experimental result reproducibility}
    \item[] Question: Does the paper fully disclose all the information needed to reproduce the main experimental results of the paper to the extent that it affects the main claims and/or conclusions of the paper (regardless of whether the code and data are provided or not)?
    \item[] Answer: \answerYes{} 
    \item[] Justification: Appendix~\ref{sec:implementation_details} provides detailed experimental information to ensure the high reproducibility of this work.
    \item[] Guidelines:
    \begin{itemize}
        \item The answer \answerNA{} means that the paper does not include experiments.
        \item If the paper includes experiments, a \answerNo{} answer to this question will not be perceived well by the reviewers: Making the paper reproducible is important, regardless of whether the code and data are provided or not.
        \item If the contribution is a dataset and\slash or model, the authors should describe the steps taken to make their results reproducible or verifiable. 
        \item Depending on the contribution, reproducibility can be accomplished in various ways. For example, if the contribution is a novel architecture, describing the architecture fully might suffice, or if the contribution is a specific model and empirical evaluation, it may be necessary to either make it possible for others to replicate the model with the same dataset, or provide access to the model. In general. releasing code and data is often one good way to accomplish this, but reproducibility can also be provided via detailed instructions for how to replicate the results, access to a hosted model (e.g., in the case of a large language model), releasing of a model checkpoint, or other means that are appropriate to the research performed.
        \item While NeurIPS does not require releasing code, the conference does require all submissions to provide some reasonable avenue for reproducibility, which may depend on the nature of the contribution. For example
        \begin{enumerate}
            \item If the contribution is primarily a new algorithm, the paper should make it clear how to reproduce that algorithm.
            \item If the contribution is primarily a new model architecture, the paper should describe the architecture clearly and fully.
            \item If the contribution is a new model (e.g., a large language model), then there should either be a way to access this model for reproducing the results or a way to reproduce the model (e.g., with an open-source dataset or instructions for how to construct the dataset).
            \item We recognize that reproducibility may be tricky in some cases, in which case authors are welcome to describe the particular way they provide for reproducibility. In the case of closed-source models, it may be that access to the model is limited in some way (e.g., to registered users), but it should be possible for other researchers to have some path to reproducing or verifying the results.
        \end{enumerate}
    \end{itemize}

\item {\bf Open access to data and code}
    \item[] Question: Does the paper provide open access to the data and code, with sufficient instructions to faithfully reproduce the main experimental results, as described in supplemental material?
    \item[] Answer: \answerYes{} 
    \item[] Justification: The data is publicly available with appropriate references provided in the main text, and the code is publicly available in our GitHub repository (\url{https://github.com/wantbook-book/NINJA}) with sufficient instructions.
    \item[] Guidelines:
    \begin{itemize}
        \item The answer \answerNA{} means that paper does not include experiments requiring code.
        \item Please see the NeurIPS code and data submission guidelines (\url{https://neurips.cc/public/guides/CodeSubmissionPolicy}) for more details.
        \item While we encourage the release of code and data, we understand that this might not be possible, so \answerNo{} is an acceptable answer. Papers cannot be rejected simply for not including code, unless this is central to the contribution (e.g., for a new open-source benchmark).
        \item The instructions should contain the exact command and environment needed to run to reproduce the results. See the NeurIPS code and data submission guidelines (\url{https://neurips.cc/public/guides/CodeSubmissionPolicy}) for more details.
        \item The authors should provide instructions on data access and preparation, including how to access the raw data, preprocessed data, intermediate data, and generated data, etc.
        \item The authors should provide scripts to reproduce all experimental results for the new proposed method and baselines. If only a subset of experiments are reproducible, they should state which ones are omitted from the script and why.
        \item At submission time, to preserve anonymity, the authors should release anonymized versions (if applicable).
        \item Providing as much information as possible in supplemental material (appended to the paper) is recommended, but including URLs to data and code is permitted.
    \end{itemize}

\item {\bf Experimental setting/details}
    \item[] Question: Does the paper specify all the training and test details (e.g., data splits, hyperparameters, how they were chosen, type of optimizer) necessary to understand the results?
    \item[] Answer: \answerYes{} 
    \item[] Justification: Appendix~\ref{sec:implementation_details} outlines all training and testing details necessary for understanding the results.
    \item[] Guidelines:
    \begin{itemize}
        \item The answer \answerNA{} means that the paper does not include experiments.
        \item The experimental setting should be presented in the core of the paper to a level of detail that is necessary to appreciate the results and make sense of them.
        \item The full details can be provided either with the code, in appendix, or as supplemental material.
    \end{itemize}

\item {\bf Experiment statistical significance}
    \item[] Question: Does the paper report error bars suitably and correctly defined or other appropriate information about the statistical significance of the experiments?
    \item[] Answer: \answerYes{} 
    \item[] Justification: All evaluation results reported in Section~\ref{sec:experiments} are computed over multiple independent runs, except for Tab.~\ref{tab:downstream_resolution_hints}. Since this experiment requires a large number of Claude API calls, we run it once and report the resulting values.
    
    \item[] Guidelines:
    \begin{itemize}
        \item The answer \answerNA{} means that the paper does not include experiments.
        \item The authors should answer \answerYes{} if the results are accompanied by error bars, confidence intervals, or statistical significance tests, at least for the experiments that support the main claims of the paper.
        \item The factors of variability that the error bars are capturing should be clearly stated (for example, train/test split, initialization, random drawing of some parameter, or overall run with given experimental conditions).
        \item The method for calculating the error bars should be explained (closed form formula, call to a library function, bootstrap, etc.)
        \item The assumptions made should be given (e.g., Normally distributed errors).
        \item It should be clear whether the error bar is the standard deviation or the standard error of the mean.
        \item It is OK to report 1-sigma error bars, but one should state it. The authors should preferably report a 2-sigma error bar than state that they have a 96\% CI, if the hypothesis of Normality of errors is not verified.
        \item For asymmetric distributions, the authors should be careful not to show in tables or figures symmetric error bars that would yield results that are out of range (e.g., negative error rates).
        \item If error bars are reported in tables or plots, the authors should explain in the text how they were calculated and reference the corresponding figures or tables in the text.
    \end{itemize}

\item {\bf Experiments compute resources}
    \item[] Question: For each experiment, does the paper provide sufficient information on the computer resources (type of compute workers, memory, time of execution) needed to reproduce the experiments?
    \item[] Answer: \answerYes{} 
    \item[] Justification: Appendix~\ref{sec:implementation_details} provides a detailed description of the computer resources utilized in this work.
    \item[] Guidelines:
    \begin{itemize}
        \item The answer \answerNA{} means that the paper does not include experiments.
        \item The paper should indicate the type of compute workers CPU or GPU, internal cluster, or cloud provider, including relevant memory and storage.
        \item The paper should provide the amount of compute required for each of the individual experimental runs as well as estimate the total compute. 
        \item The paper should disclose whether the full research project required more compute than the experiments reported in the paper (e.g., preliminary or failed experiments that didn't make it into the paper). 
    \end{itemize}
    
\item {\bf Code of ethics}
    \item[] Question: Does the research conducted in the paper conform, in every respect, with the NeurIPS Code of Ethics \url{https://neurips.cc/public/EthicsGuidelines}?
    \item[] Answer: \answerYes{} 
    \item[] Justification: The work adheres fully to the NeurIPS Code of Ethics.
    \item[] Guidelines:
    \begin{itemize}
        \item The answer \answerNA{} means that the authors have not reviewed the NeurIPS Code of Ethics.
        \item If the authors answer \answerNo, they should explain the special circumstances that require a deviation from the Code of Ethics.
        \item The authors should make sure to preserve anonymity (e.g., if there is a special consideration due to laws or regulations in their jurisdiction).
    \end{itemize}

\item {\bf Broader impacts}
    \item[] Question: Does the paper discuss both potential positive societal impacts and negative societal impacts of the work performed?
    \item[] Answer: \answerYes{} 
    \item[] Justification: Appendix~\ref{app:broader_impact} provides a detailed discussion of the potential societal impacts.
    \item[] Guidelines:
    \begin{itemize}
        \item The answer \answerNA{} means that there is no societal impact of the work performed.
        \item If the authors answer \answerNA{} or \answerNo, they should explain why their work has no societal impact or why the paper does not address societal impact.
        \item Examples of negative societal impacts include potential malicious or unintended uses (e.g., disinformation, generating fake profiles, surveillance), fairness considerations (e.g., deployment of technologies that could make decisions that unfairly impact specific groups), privacy considerations, and security considerations.
        \item The conference expects that many papers will be foundational research and not tied to particular applications, let alone deployments. However, if there is a direct path to any negative applications, the authors should point it out. For example, it is legitimate to point out that an improvement in the quality of generative models could be used to generate Deepfakes for disinformation. On the other hand, it is not needed to point out that a generic algorithm for optimizing neural networks could enable people to train models that generate Deepfakes faster.
        \item The authors should consider possible harms that could arise when the technology is being used as intended and functioning correctly, harms that could arise when the technology is being used as intended but gives incorrect results, and harms following from (intentional or unintentional) misuse of the technology.
        \item If there are negative societal impacts, the authors could also discuss possible mitigation strategies (e.g., gated release of models, providing defenses in addition to attacks, mechanisms for monitoring misuse, mechanisms to monitor how a system learns from feedback over time, improving the efficiency and accessibility of ML).
    \end{itemize}
    
\item {\bf Safeguards}
    \item[] Question: Does the paper describe safeguards that have been put in place for responsible release of data or models that have a high risk for misuse (e.g., pre-trained language models, image generators, or scraped datasets)?
    \item[] Answer: \answerNA{} 
    \item[] Justification: This work does not release models or datasets with a high risk of misuse. As discussed in Appendix F, we recommend applying the method only to authorized repositories to mitigate potential risks.
    \item[] Guidelines:
    \begin{itemize}
        \item The answer \answerNA{} means that the paper poses no such risks.
        \item Released models that have a high risk for misuse or dual-use should be released with necessary safeguards to allow for controlled use of the model, for example by requiring that users adhere to usage guidelines or restrictions to access the model or implementing safety filters. 
        \item Datasets that have been scraped from the Internet could pose safety risks. The authors should describe how they avoided releasing unsafe images.
        \item We recognize that providing effective safeguards is challenging, and many papers do not require this, but we encourage authors to take this into account and make a best faith effort.
    \end{itemize}

\item {\bf Licenses for existing assets}
    \item[] Question: Are the creators or original owners of assets (e.g., code, data, models), used in the paper, properly credited and are the license and terms of use explicitly mentioned and properly respected?
    \item[] Answer: \answerYes{} 
    \item[] Justification:  The assets utilized in this paper are appropriately referenced and accessible to the public.
    \item[] Guidelines:
    \begin{itemize}
        \item The answer \answerNA{} means that the paper does not use existing assets.
        \item The authors should cite the original paper that produced the code package or dataset.
        \item The authors should state which version of the asset is used and, if possible, include a URL.
        \item The name of the license (e.g., CC-BY 4.0) should be included for each asset.
        \item For scraped data from a particular source (e.g., website), the copyright and terms of service of that source should be provided.
        \item If assets are released, the license, copyright information, and terms of use in the package should be provided. For popular datasets, \url{paperswithcode.com/datasets} has curated licenses for some datasets. Their licensing guide can help determine the license of a dataset.
        \item For existing datasets that are re-packaged, both the original license and the license of the derived asset (if it has changed) should be provided.
        \item If this information is not available online, the authors are encouraged to reach out to the asset's creators.
    \end{itemize}

\item {\bf New assets}
    \item[] Question: Are new assets introduced in the paper well documented and is the documentation provided alongside the assets?
    \item[] Answer: \answerYes{} 
    \item[] Justification: Our code is publicly available in our GitHub repository (\url{https://github.com/wantbook-book/NINJA}) and includes detailed instructions for use.
    \item[] Guidelines:
    \begin{itemize}
        \item The answer \answerNA{} means that the paper does not release new assets.
        \item Researchers should communicate the details of the dataset\slash code\slash model as part of their submissions via structured templates. This includes details about training, license, limitations, etc. 
        \item The paper should discuss whether and how consent was obtained from people whose asset is used.
        \item At submission time, remember to anonymize your assets (if applicable). You can either create an anonymized URL or include an anonymized zip file.
    \end{itemize}

\item {\bf Crowdsourcing and research with human subjects}
    \item[] Question: For crowdsourcing experiments and research with human subjects, does the paper include the full text of instructions given to participants and screenshots, if applicable, as well as details about compensation (if any)? 
    \item[] Answer: \answerNA{} 
    \item[] Justification: This work does not involve Crowdsourcing or research with human subjects.
    \item[] Guidelines:
    \begin{itemize}
        \item The answer \answerNA{} means that the paper does not involve crowdsourcing nor research with human subjects.
        \item Including this information in the supplemental material is fine, but if the main contribution of the paper involves human subjects, then as much detail as possible should be included in the main paper. 
        \item According to the NeurIPS Code of Ethics, workers involved in data collection, curation, or other labor should be paid at least the minimum wage in the country of the data collector. 
    \end{itemize}

\item {\bf Institutional review board (IRB) approvals or equivalent for research with human subjects}
    \item[] Question: Does the paper describe potential risks incurred by study participants, whether such risks were disclosed to the subjects, and whether Institutional Review Board (IRB) approvals (or an equivalent approval/review based on the requirements of your country or institution) were obtained?
    \item[] Answer: \answerNA{} 
    \item[] Justification: This work does not involve research with human subjects.
    \item[] Guidelines:
    \begin{itemize}
        \item The answer \answerNA{} means that the paper does not involve crowdsourcing nor research with human subjects.
        \item Depending on the country in which research is conducted, IRB approval (or equivalent) may be required for any human subjects research. If you obtained IRB approval, you should clearly state this in the paper. 
        \item We recognize that the procedures for this may vary significantly between institutions and locations, and we expect authors to adhere to the NeurIPS Code of Ethics and the guidelines for their institution. 
        \item For initial submissions, do not include any information that would break anonymity (if applicable), such as the institution conducting the review.
    \end{itemize}

\item {\bf Declaration of LLM usage}
    \item[] Question: Does the paper describe the usage of LLMs if it is an important, original, or non-standard component of the core methods in this research? Note that if the LLM is used only for writing, editing, or formatting purposes and does \emph{not} impact the core methodology, scientific rigor, or originality of the research, declaration is not required.
    \item[] Answer: \answerNo{} 
    \item[] Justification: We use it solely for grammar and phrasing refinement to improve the clarity of the paper.
    \item[] Guidelines:
    \begin{itemize}
        \item The answer \answerNA{} means that the core method development in this research does not involve LLMs as any important, original, or non-standard components.
        \item Please refer to our LLM policy in the NeurIPS handbook for what should or should not be described.
    \end{itemize}

\end{enumerate}

\end{document}